%% file: main.tex
\documentclass[journal]{IEEEtran}
\ifCLASSINFOpdf
\else
\fi
\usepackage{amsmath}
\usepackage{graphicx}
\usepackage{multirow}

\usepackage{tikz}
\usetikzlibrary{spy}

\input{commands.tex}

\begin{document}
%
\title{Deep Graph-Convolutional Image Denoising}
%
%
%

\author{Diego Valsesia,
        Giulia Fracastoro,
        Enrico Magli
\thanks{The authors are with Politecnico di Torino -- Department of Electronics and Telecommunications, Italy. email: \{name.surname\}@polito.it. This research has been partially funded by the SmartData@PoliTO center for Big Data and Machine Learning technologies. We thank Nvidia for donating a Quadro P6000 GPU.} }
\maketitle

\begin{abstract}
Non-local self-similarity is well-known to be an effective prior for the image denoising problem. However, little work has been done to incorporate it in convolutional neural networks, which surpass non-local model-based methods despite only exploiting local information. In this paper, we propose a novel end-to-end trainable neural network architecture employing layers based on graph convolution operations, thereby creating neurons with non-local receptive fields. The graph convolution operation generalizes the classic convolution to arbitrary graphs. In this work, the graph is dynamically computed from similarities among the hidden features of the network, so that the powerful representation learning capabilities of the network are exploited to uncover self-similar patterns. We introduce a lightweight Edge-Conditioned Convolution which addresses vanishing gradient and over-parameterization issues of this particular graph convolution. Extensive experiments show state-of-the-art performance with improved qualitative and quantitative results on both synthetic Gaussian noise and real noise.
\end{abstract}


%
\IEEEpeerreviewmaketitle

\section{Introduction}

Denoising is a staple among image processing problems and its importance cannot be overstated. Despite decades of work and countless methods, it still remains an active research topic because its purpose goes far beyond generating visually pleasing pictures. Denoising is fundamental to enhance the performance of higher-level computer vision tasks such as classification, segmentation or object recognition, and is a building block in the solution to various problems \cite{lukavs2006digital,valsesia2015compressed,romano2017little,sun2019block}. The recent successes achieved by convolutional neural networks (CNNs) extended to this problem as well and have brought a new generation of learning-based methods that is redefining the state of the art. However, it is important to learn the lessons of past research on the topic and integrate them with the new deep learning techniques. In particular, classic denoising methods such as BM3D \cite{dabov2007image} showed the importance of exploiting non-local self-similar patterns. However, the convolution operation underpinning all CNNs architectures \cite{zhang2017beyond,mao2016image,tai2017memnet,bae2017beyond} is unable to capture such patterns because of the locality of the convolution kernels. Only very recently, some works started addressing the integration of non-local information into CNNs \cite{cruz2018nonlocality,lefkimmiatis2018universal,plotz2018neural,liu2018non}.

This paper presents a denoising neural network, called GCDN, where the convolution operation is generalized by means of graph convolution, which is used to create layers with hidden neurons having non-local receptive fields that successfully capture self-similar information. Graph convolution is a generalization of the traditional convolution operation when the data are represented as sitting over the vertices of a graph. In this work, every pixel is a vertex and the edges in the graph are dynamically computed from the similarities in the feature space of the hidden layers of the network. This allows us to exploit the powerful representational features of neural networks to discover and use latent self-similarities. With respect to other CNNs integrating non-local information for the denoising task, the proposed approach has several advantages: i) it creates an adaptive receptive field for the pixels in the hidden layers by dynamically computing a nearest-neighbor graph from the latent features; ii) it creates dynamic non-local filters where feature vectors that may be spatially distant but close in a latent vector space are aggregated with weights that depend on the features themselves; iii) the aggregation weights are estimated by a fully-learned operation, implemented as a subnetwork, instead of a predefined parameterized operation, allowing more generality and adaptability. Starting from the Edge-Conditioned Convolution (ECC) definition of graph convolution, we propose several improvements to address stability, over-parameterization and vanishing gradient issues. Finally, we also propose a novel neural network architecture which draws from an analogy with an unrolled regularized optimization method.

A preliminary version of this work appeared in \cite{ValsesiaICIP19}. There are several differences with the work in this paper. The architecture of the network is improved by drawing an analogy with proximal gradient descent methods, and it is significantly deeper. Moreover, we propose several solutions to address the ECC overparameterization and computational issues. Finally, we also present an in-depth analysis of the network behavior and greatly extended experimental results.

This paper is structured as follows. Sec. \ref{sec:bkg} provides some background material on graph-convolutional neural networks and state-of-the-art denoising approaches. Sec. \ref{sec:method} describes the proposed method. Sec. \ref{sec:experiments} analyzes the characteristics of the proposed method and experimentally compares it with state-of-the-art approaches. Finally, Sec. \ref{sec:conclusions} draws some conclusions.

\section{Related work} \label{sec:bkg}

\subsection{Graph neural networks}
Inspired by the overwhelming success of deep neural networks in computer vision, a significant research effort has recently been made in order to develop deep learning methods for data that naturally lie on irregular domains. One case is when the data domain can be structured as a graph and the data are defined as vectors on the nodes of this graph. Extending CNNs from signals with a regular structure, such as images and video, to graph-structured signals is not straightforward, since even simple operations such as shifts are undefined over graphs. 

One of the major challenges in this field is defining a convolution-like operation for this kind of data. Convolution has a key role in classical CNNs, thanks to its properties of locality, stationarity, compositionality, which well match prior knowledge on many kinds of data and thus allow effective weight reuse. For this reason, defining an operation with similar characteristics for graph-structured data is of primary importance in order to obtain effective graph neural networks. The literature has identified two main classes of approaches to tackle this problem, namely spectral or spatial. In the former case \cite{henaff2015deep,defferrard2016convolutional,kipf2016semi}, the convolution is defined in the spectral domain through the graph Fourier transform \cite{shuman2013emerging}. Fast polynomial approximations \cite{defferrard2016convolutional} have been proposed in order to obtain an efficient implementation of this operation. Graph-convolutional neural networks (GCNN) with this convolution operator have been successfully applied in problems of semi-supervised node classification and link prediction \cite{kipf2016semi,schlichtkrull2018modeling}. The main drawback of these methods is that the graph is supposed to be fixed and it is not clear how to handle the cases where the structure varies. The latter class of approaches overcomes this issue by defining the convolution operator in the spatial domain \cite{simonovsky2017dynamic,wang2018dynamic,xu2018powerful,monti2017geometric,verma2018feastnet,valsesia2019learning}. In this case, the convolution is performed by local aggregations, i.e. a weighted combination of the signal values over neighboring nodes. Since in this case the operation is defined at a neighborhood level, the convolution remains well-defined even when the graph structure varies. Many of the spatial approaches present in the literature \cite{xu2018powerful,monti2017geometric,verma2018feastnet} perform local aggregations with scalar weights. Instead, \cite{simonovsky2017dynamic} proposes to weight the contributions of the neighbors using edge-dependent matrices. This makes the convolution a more general function, increasing its descriptive power. For this reason, in this paper we employ the convolution operator proposed in \cite{simonovsky2017dynamic}. However, in order to obtain an efficient operation, we introduce several approximations that reduce its computation complexity, memory occupation, and mitigate vanishing gradient issues that arise when trying to build very deep architectures.

\subsection{Image denoising}
The literature on image denoising is vast, as it is one of most classic problems in image processing. Focusing on the recent developments, we can broadly define two categories of methods: model-based approaches and learning-based approaches. 

Model-based approaches traditionally focused on defining hand-crafted priors to carefully capture the salient features of natural images. Early works in this category include total variation minimization \cite{rudin1992nonlinear}, and bilateral filtering \cite{tomasi1998bilateral}. Non-local means \cite{buades2005non} introduced the idea of non-local averaging according to the similarity of local neighborhood. The popular BM3D \cite{dabov2007image} expanded on the idea by collaborative filtering of the matched patches. WNNM \cite{gu2014weighted} used nuclear norm minimization to enforce a low-rank prior. Finally, some works recently introduced graph-based regularizers \cite{pang2017graph} to enforce a measure of smoothness of the signal across the edges of a graph of patch or pixel similiarities.   
Many of the most successful model-based approaches are non-local, i.e., they exploit the concept of self-similarity among structures in the image beyond the local neighborhood.

\begin{figure*}
    \centering
    \includegraphics[width=\textwidth]{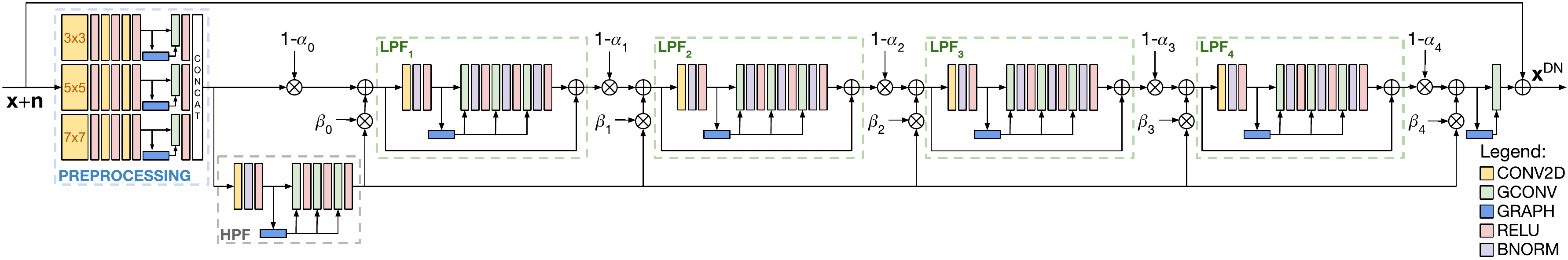}
    \caption{GCDN architecture.}
    \label{fig:net}
\end{figure*}
Learning-based approaches use training data to learn a model for natural images. The popular K-SVD algorithm \cite{elad2006image} learns a dictionary in which natural patches have a sparse representation, and therefore casts image denoising as a sparse coding problem on this learned dictionary. The TNRD method \cite{chen2016trainable} uses a nonlinear reaction diffusion model with trainable filters. An early work with neural networks \cite{burger2012image} used a multilayer perceptron discriminatively trained on synthetic Gaussian noise and showed significant improvements over model-based methods. More recently, CNNs have achieved remarkable performance. Zhang et al. \cite{zhang2017beyond} showed that the residual structure and the use of batch normalization \cite{ioffe2015batchnorm} in their DnCNN greatly helps the denoising task. Following the DnCNN, many other architectures have been proposed, such as RED \cite{mao2016image}, MemNet \cite{tai2017memnet} and a CNN working on wavelet coefficients \cite{bae2017beyond}. However, those CNN-based methods are limited by the local nature of the convolution operation, which is unable to increase the receptive field of a neuron-pixel to model non-local image features. This means that CNNs are unable to exploit the self-similar patterns that were proven to be highly successful in model-based methods. Very recently, a few works started addressing this issue by trying to incorporate non-local information in a CNN. NN3D \cite{cruz2018nonlocality} uses a global post-processing stage based on a non-local filter after the output of a denoising CNN. This stage performs block matching and filtering over the whole image denoised by the CNN. This is clearly suboptimal as the non-local information does not contribute to the training of the CNN. 
UNLNet \cite{lefkimmiatis2018universal} introduces a trainable non-local layer which collaboratively filters image blocks. However, performance is limited by the selection of matching blocks from the noisy input image instead of the feature space, and ultimately UNLNet does not improve over the performance of the simpler DnCNN. N$^3$Net \cite{plotz2018neural} introduces a continuous nearest-neighbor relaxation to create a non-local layer. Finally, NLRN \cite{liu2018non} proposes a non-local module that uses the distances among hidden feature vectors of a search window around the pixel of interest to aggregate such vectors and return the output features of the pixel. However, there are significant differences with respect to the work in this paper. First, they use all the pixels in the search window instead of only a number of nearest neighbors, which means that their receptive field cannot dynamically adapt to the content of the image. Then, while in both works the feature aggregation weights are dynamically computed from the features themselves, NLRN uses an explicitly-parameterized function with learnable parameters, in contrast to this work where the function is fully learned as a dedicated sub-network. These choices increase the adaptivity of the proposed non-local operations, which result in better performance around edges.

\section{Proposed denoiser}\label{sec:method}

\subsection{Overview} \label{sec:overview}

An overview of the proposed graph-convolutional denoiser network (GCDN) can be seen in Fig. \ref{fig:net}. The structure will be explained more in detail in Sec. \ref{sec:analogy} where an analogy is drawn between unrolled proximal gradient descent with a graph total variation regularizer and the proposed network architecture. At a first glance, the network has a global input-output residual connection whereby the network learns to estimate the noise rather than successively clean the image. This has been shown \cite{zhang2017beyond} to improve training convergence for the denoising problem. 

The main feature of the proposed network is the use of graph-convolutional layers where the graphs are dynamically computed from the feature space. The graph-convolutional layer, described in Sec. \ref{sec:gconv_layer}, creates a non-local receptive field for each pixel-neuron, so that pixels that are spatially distant but similar in the feature space created by the network can be merged. 

An important block of the proposed network is the preprocessing stage at the input. It can be noticed that the first layers of the network are classic 2D convolutions rather than graph convolutions. This is done to create an embedding over a receptive field larger than a single pixel and stabilize the graph construction operation, which would otherwise be affected by the input noise. The preprocessing stage has three parallel branches that operate on multiple scales, in a fashion similar to the architectures in \cite{szegedy2015going} and \cite{divakar2017image}. The multiscale features are extracted by a sequence of three convolutional layers with filters of size $3\times 3$, $5\times 5$, and $7\times 7$, depending on the branch. After a final graph-convolutional layer, the features are concatenated. 

The remaining network layers are grouped into an HPF block and multiple LPF blocks, named after the analogy with highpass and lowpass graph filters described in Sec. \ref{sec:analogy}. These blocks have an initial $3 \times 3$ convolutional layer followed by three graph-convolutional layers sharing the same graph constructed from the output of the convolutional layer. All layers are interleaved by Batch Normalization operations \cite{ioffe2015batchnorm} and leaky ReLU nonlinearities. Notice that the LPF blocks have themselves a residual connection to help backpropagation, as in ResNet architectures \cite{he2016deep}. The final layer is a graph-convolutional layer mapping from feature space to the image space.

\subsection{Graph-convolutional layer} \label{sec:gconv_layer}
\label{sec:graph_conv}
\begin{figure}
    \centering
    \includegraphics[width=0.23\textwidth]{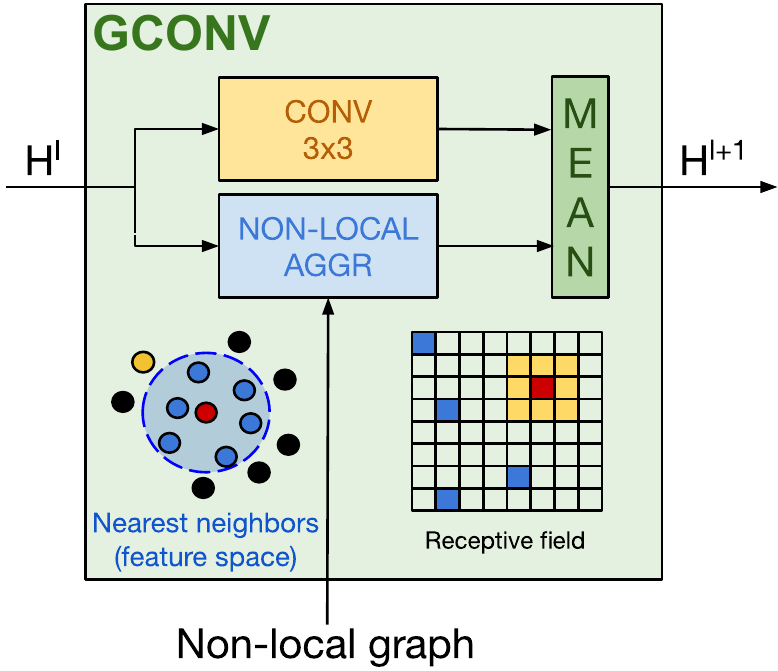}
    \caption{Graph-convolutional layer. The operation has a receptive field with a local component ($3\times 3$ 2D convolution) and a non-local component (pixels selected as nearest neighbors in the feature space).}
    \label{fig:gconv}
\end{figure}

The operation performed by the graph-convolutional layer is summarized in Fig. \ref{fig:gconv}. The two inputs to the graph-convolutional layer are the feature vectors $\Hb^l \in \RR^{F^l \times N}$ associated to the $N$ image pixels at layer $l$ and the adjacency matrix of a graph connecting image pixels. In this work, the graph is constructed as a $K$-nearest neighbor graph in the feature space. For each pixel, the Euclidean distances between its feature vector and the feature vectors of pixels inside a search window are computed and an edge is drawn between the pixel and the $K$ pixels with smallest distance. Using this method, we obtain a $K$-regular graph $\G^l(\V,\E^l)$, where $\V$ is the set of vertices with $|\V|=N$ and $\E^l\subseteq \mathcal{V}\times\mathcal{V}$ is the set of edges. We also assume that the edges of $\G^l$ are labeled, i.e. there exists a function $\mathcal{L}:\mathcal{E}^l\to \RR^{F^l}$ that assigns a label to each edge. In this work, we define the edge labeling function as the difference between the two feature vectors, i.e. $\mathcal{L}(i,j)=\Hb_j^l-\Hb_i^l=\mathbf{d}^{l,j\to i}$.  A classic $3 \times 3$ local convolution processes the local neighborhood to provide its estimate of the output feature vector for the current pixel, while the feature vectors of the non-local pixels connected by the graph are aggregated by means of the edge-conditioned convolution (ECC) \cite{simonovsky2017dynamic}. Notice that the 8 local neighbors of the pixel are excluded from graph construction as they are already used by the local convolution.
The non-local aggregation is computed as:
\begin{align}
\label{eq:ecc}
\Hb_i^{l+1,\mathrm{NL}}  &=  \sum_{j\in\Sx_i^l} \gamma^{l,j\to i} \frac{\fun_{\wb^l}^l\left(\mathbf{d}^{l,j\to i}\right)\Hb_j^l}{|\Sx_i^l|}\nonumber\\&= \sum_{j\in\Sx_i^l} \gamma^{l,j\to i} \frac{\bm{\Theta}^{l,j\to i}\Hb_j^l}{|\Sx_i^l|},
\end{align}
where $\fun_{\wb^l}^l:\RR^{F^l}\to\RR^{F^{l+1}\times F^{l}}$ is a fully-connected network that takes as input the edge labels and outputs the corresponding weight matrix $\bm{\Theta}^{l,j\to i} = \fun_{\wb^l}^l\left(\mathcal{L}(i,j)\right) \in\RR^{F^{l+1}\times F^{l}}$, $\wb^l$ are the weights parameterizing network $\fun^l$, and $\Sx_i^l$ is the set of neighbors of node $i$ in the graph $\G^l$. The scalar $\gamma^{j\to i}$ is an edge-attention term computed as:
\begin{align} \label{eq:edge_attention}
    \gamma^{l,j\to i} = \exp\left( - \Vert \mathbf{d}^{l,j\to i} \Vert^2_2 / \delta \right)
\end{align}
where $\delta$ is a cross-validated hyper-parameter. This term is reminiscent of the edge attention mechanism from the graph neural network literature \cite{gong2019exploiting} and it serves the purpose of stabilizing training by underweighting the edges that connect nodes with distant feature vectors. Note that this term could, in principle, be learned by the $\fun$ network but we found that decoupling it and making it explicitly dependent on feature distances in an exponential way, accelerated and stabilized training. Also notice that in Sec. \ref{sec:experiments} we show that this term alone, i.e. without weight matrices $\bm{\Theta}$, is not powerful enough to reach good performance. Moreover, it is worth mentioning that the edge weights $\bm{\Theta}$ and the edge-attention term $\gamma$ depend only on the edge labels. This means that two pairs of nodes with the same edge labels will have the same weights, resulting in a behaviour similar to weight sharing in classical CNNs.

Finally, we combine the feature vector estimated by the non-local aggregation with the one produced by the local convolution to provide the output features as follows
\[
\Hb_i^{l+1}=\frac{\Hb_i^{l+1,\mathrm{NL}}+\Hb_i^{l+1,\mathrm{L}}}{2}+\mathbf{b}^l,
\]
where $\Hb_i^{l+1,\mathrm{L}}$ is the output of the $3\times 3$ local convolution for the node $i$ and $\mathbf{b}^l\in\RR^{F^l}$ is the bias.

The advantages of the ECC with respect to other definitions of graph convolution are trifold: i) the edge weights depend on the edge label, ii) it allows to compute an affine transformation along every edge, and iii) the edge weight function is highly general since it does not have a predefined structure. By making the edge weights depend on the input features, the ECC implements an adaptive filter which can be more complex than the non-adaptive local filters. Moreover, the second advantage is due to the fact that $\bm{\Theta}^{l,j\to i}$ is an edge-dependent matrix, making the convolution operation more general than other non-local aggregation methods using scalar edge weights. Among such methods we can find GCN \cite{kipf2016semi}, GIN \cite{xu2018powerful}, MoNet \cite{monti2017geometric}, and FeastNet \cite{verma2018feastnet}.  Finally, the $\fun$ function is a general function which can be learned to be the optimal one for the denoising task by the function approximation capability of the subnetwork implementing it. This is in contrast with other methods where the function predicting the edge weights is fixed with some learnable parameters. For example, FeastNet \cite{verma2018feastnet} employs scalar edge weights computed using the following function
\[
f(\Hb_i^{l},\Hb_j^{l})\propto \exp \left(\mathbf{u}^T\Hb_i^{l}+\mathbf{v}^T\Hb_j^{l}+c\right),
\]
where $\mathbf{u},\mathbf{v}\in\RR^{F^l}$ and $c\in\RR$ are learnable parameters. Instead, MoNet \cite{monti2017geometric} employs a Gaussian kernel as follows
\[
f(\Hb_i^{l},\Hb_j^{l})=\exp\left(-\frac{1}{2} (\mathbf{d}^{l,j\to i}-\bm{\mu})^T\bm{\Sigma}^{-1}(\mathbf{d}^{l,j\to i}-\bm{\mu})\right),
\]
where $\bm{\Sigma}\in\RR^{F^l\times F^l}$ and $\bm{\mu}\in\RR^{F^l}$ are learnable parameters. Also NLRN \cite{liu2018non} uses a Gaussian kernel to perform non-local aggregations. We can consider this operation as a graph convolution where each pixel is connected to all the other pixels in its search window and the edge weights are defined as follows
\[
f(\Hb_i^{l},\Hb_j^{l})=\frac{\exp\left(\Hb_i^{lT}\Wb_{\theta}^T\Wb_{\phi}\Hb_j^{l}\right)}{\sum_{j\in\mathcal{S}_i}\exp\left(\Hb_i^{lT}\Wb_{\theta}^T\Wb_{\phi}\Hb_j^{l}\right)}\Wb_g,
\]
where $\Wb_{\theta},\Wb_{\phi}\in\RR^{t\times F^l}$ and $\Wb_g\in\RR^{F^{l+1}\times F^{l}}$ are learnable parameters.

\subsection{Lightweight Edge-Conditioned Convolution} \label{sec:ecc}
As seen in the previous section, the function $\fun$ has a key role in the ECC because it defines the weights for the neighborhood aggregation. In the original definition of ECC \cite{simonovsky2017dynamic}, the function $\fun$ is implemented as a two-layer fully connected network. This definition raises some relevant issues. In the following, we will describe in detail these issues and present two possible solutions.

\subsubsection{Circulant approximation of dense layer}

\begin{figure}
    \centering
    \includegraphics[width=0.3\textwidth]{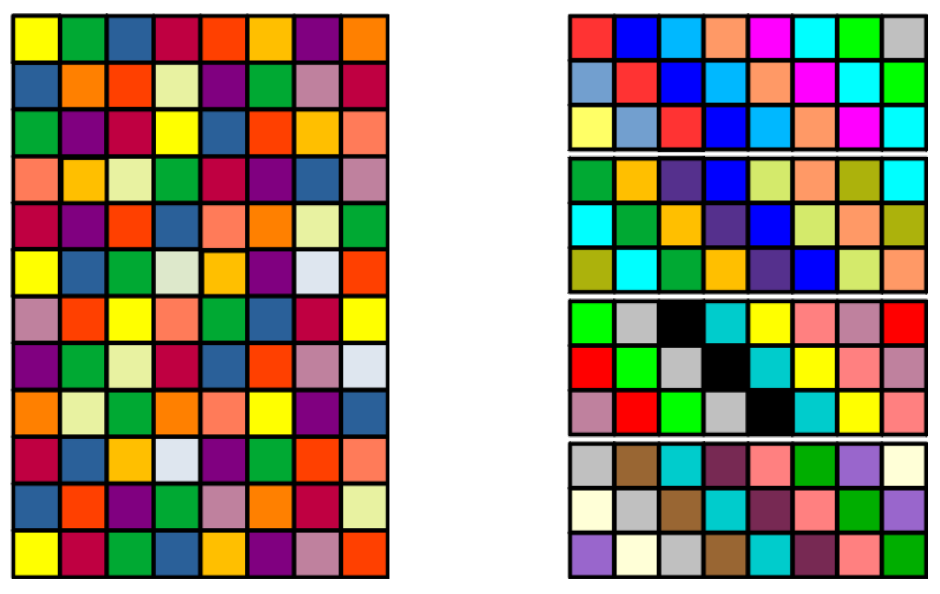}
    \caption{Circulant approximation of a fully-connected layer.}
    \label{fig:circ}
\end{figure}

The first issue is related to the risk of over-parameterization. The dimension of the input of the $\fun$ network is $F^l$, while the dimension of its output is $F^{l+1}\times F^{l}$. This means that the number of weights of the network depends cubically on the number of features. Therefore, the number of parameters quickly becomes excessively large, resulting in vanishing gradients or overfitting. 


To address the over-parameterization problem we propose to use a partially-structured matrix for the last layer, instead of an unstructured one. We impose that this matrix is composed of multiple stacked partial circulant matrices, i.e., matrices where only a few shifted versions of the first row are used instead of all the possible ones of the full square matrix. Fig. \ref{fig:circ} shows the structure of the approximated matrix. Using this approximation, the only free parameters are in the first row of each partial circulant matrix. If only $m$ shifts per partial circulant matrix are allowed, we reduce the number of parameters by a factor $m$. Thus, if the unstructured dense matrix has $F^l F^{l+1} \times F^l$ parameters, with the proposed approximation the number of parameters drops to  $\frac{F^l F^{l+1}}{m}\times F^l$. Similar approaches to approximate fully connected layers have already been studied in the literature \cite{cheng2015exploration,wu2016compression}. In particular, \cite{cheng2015exploration} shows that imposing a partial circulant structure does not significantly impact the final performance in a classification problem. Indeed, there are connections with results stating that random partial circulant matrices implement stable embeddings almost as well as fully random matrices \cite{hinrichs2011johnson,valsesia2017user,valsesia2017binary}.

\subsubsection{Low-rank node aggregation}

The second issue related to the $\fun$ network regards memory occupation and computations. In order to perform the ECC operation, we have to compute a weight matrix $\bm{\Theta}^{l,j\to i}$ for each edge $j$ of every neighborhood $\Nc_i$ of every image in the batch. If we consider a $K$-regular graph and a batch of $B$ images with $N$ pixels each, the memory occupation needed to store all the matrices $\bm{\Theta}^{l,j\to i}$ as single-precision floating point tensors is equal to $B\times N\times K\times F^{l+1}\times F^l \times 4$ bytes and this quantity can easily become unmanageable. To give an idea of the required amount of memory,  let us consider an example with $B=16$, $N=1024$, $K=8$, $F^l=F^{l+1}=66$, then the memory required to store all the matrices $\bm{\Theta}^{l,j\to i}$ for only one graph-convolutional layer is around 2 GB.

\begin{figure}
    \centering
    \includegraphics[width=0.28\textwidth]{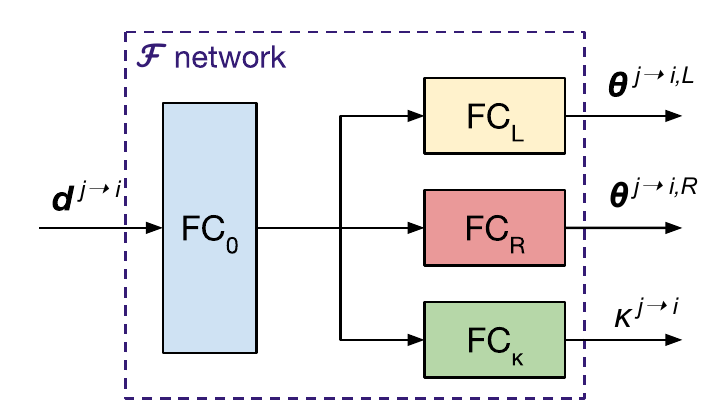}
    \caption{$\fun$ network. FC$_0$ is a fully-connected layer followed by a leaky ReLU non-linearity. The FC$_R$, FC$_L$, FC$_\kappa$ do not have any output non-linearities.}
    \label{fig:fnet}
\end{figure}

In order to solve this issue, we propose to impose a low-rank approximation for $\bm{\Theta}^{l,j\to i}$. 
Let us consider the singular value decomposition of a matrix  
\begin{align*}
\mathbf{A} = \bm{\Phi}\Lambdab \bm{\Psi}^T = \sum_{s}\lambda_s \bm{\phi}_s\bm{\psi}_s^T,
\end{align*}
where $\bm{\phi}_s$ and $\bm{\psi}_s$ are the left and right singular vectors and $\lambda_s$ the singular values. We can obtain a low-rank approximation of rank $r$ by keeping only the $r$ largest singular values and setting the others to zero. Therefore, the approximation is reduced to a sum of $r$ outer products. Inspired by this fact, we define $\bm{\Theta}^{l,j\to i}$ as follows
\begin{equation}
\bm{\Theta}^{l,j\to i}=\sum_{s=1}^r\kappa_s^{j\to i}\bm{\theta}_s^{j\to i,L} \bm{\theta}_s^{j\to i,R^T},    
\label{eq:theta_approx}
\end{equation}
where $\bm{\theta}_s^{j\to i,L}\in\mathbb{R}^{F^{l}}$, $\bm{\theta}_s^{j\to i,R}\in\mathbb{R}^{F^{l+1}}$, $\kappa_s^{j\to i}\in\mathbb{R}$ and $1\le r\le F^l$. Notice that the approximation in \eqref{eq:theta_approx} ensures that the rank is at most $r$ rather than exactly enforcing a rank-$r$ structure, because we do not impose orthogonality between $\bm{\theta}_s^{j\to i,L}$ and $\bm{\theta}_s^{j\to i,R}$, even though random initialization makes them quasi-orthogonal. Using this approximation, we can redefine the $\fun$ network in such a way that it outputs $\bm{\theta}_s^{j\to i,L},\bm{\theta}_s^{j\to i,R},\kappa_s^{j\to i}$ for $s=1,2,\dots,r$. In particular, we redefine the second layer of the $\fun$ network: instead of having a single fully connected layer that outputs the entire matrix $\bm{\Theta}^{l,j\to i}$, we have three parallel fully connected layers that separately output $\bm{\theta}_s^{j\to i,L}$, $\bm{\theta}_s^{j\to i,R}$ and $\kappa_s^{j\to i}$, as shown in Fig. \ref{fig:fnet}. The advantage of this approximation is that we only need to store $\bm{\theta}_s^{j\to i,L}$,  $\bm{\theta}_s^{j\to i,R}$ and $\kappa_s^{j\to i}$ instead of the entire matrix $\bm{\Theta}^{l,j\to i}$, drastically reducing the memory occupation to $B\times N\times K\times r(2F^l+1) \times 4$ bytes. If we consider the example presented above and set $r=10$, the memory requirement drops from 2 GB to 700 MB. Another advantage of this approximation is that it also leads to a significant reduction of the computation burden, because we never have to actually compute all the matrices $\bm{\Theta}^{l,j\to i}$. In fact, the neighborhood aggregation can be reduced as follows
\begin{align} \label{eq:lowrank_aggr}
\Hb_i^{l+1,\mathrm{NL}}&=\sum_{j\in\Sx_i^l} \gamma^{l,j\to i} \frac{\bm{\Theta}^{l,j\to i}\Hb_j^l}{|\Sx_i^l|} \nonumber\\
&=\sum_{j\in\Sx_i^l} \gamma^{l,j\to i} \frac{\sum_{s=1}^r\kappa_s^{j\to i}\bm{\theta}_s^{j\to i,L} \bm{\theta}_s^{j\to i,R^T}\Hb_j^l}{|\Sx_i^l|},
\end{align}
where the computational cost of the full operation on the first line is $O(F^l F^{l+1})$, instead the cost of the decoupled operation on the second line is $O(r(F^l+F^{l+1}))$.
Finally, this approximation also helps to reduce the number of parameters of the last layer of the $\fun$ network since the output has size $r(F^l+F^{l+1}+1)$ instead of $F^{l+1}F^l$.

When we employ the new structure of the $\fun$ network, we need to pay special attention to the weight initialization. In particular, we have to carefully define the variance of the random weight initialization of the three parallel layers to avoid scaling problems. We define $\Wb_0$ as the weight matrix of the first layer of the $\fun$ network, and $\Wb^L$, $\Wb^R$ and $\Wb^{\kappa}$ as the weight matrices of the three parallel fully connected layers. Let us suppose that $\mathbf{d}^{j\to i}_t$ has been normalized to be approximately a standard Gaussian, i.e., $\mathbf{d}^{j\to i}_t  \sim \Nc(0,1)$ for $t=1,\dots, F^l$, and that $\Wb^0$ has been initialized using Glorot initialization \cite{pmlr-v9-glorot10a}, i.e., $\Wb^0_{uv}\sim\Nc\left(0,\frac{1}{F^l}\right)$ with $u,v = 1,\dots, F^{l}$. Let us also assume that $\Wb^L_{uv}\sim\Nc(0,\sigma^2_L)$, $\Wb^R_{uv}\sim\Nc(0,\sigma^2_R)$, and $\Wb^{\kappa}_{u}\sim\Nc(0,\sigma^2_{\kappa})$. Then, we obtain
\[
\begin{split}
    \bm{\theta}_{s,u}^{j\to i,L}&\sim\Nc(0,F^l\sigma^2_L),\\
    \bm{\theta}_{s,u}^{j\to i,R}&\sim\Nc(0,F^l\sigma^2_R),\\
    \kappa_s^{j\to i}&\sim \Nc(0,F^l\sigma_{\kappa}^2),
\end{split}
\]
where $s=1,\dots,r$. Finally, considering the aggregation formula in Eq. \eqref{eq:lowrank_aggr} leads to the following result:
\begin{equation}
\label{eq:init}
\Hb_{i,u}^{l+1,\mathrm{NL}} \sim \Nc\left(0,\frac{1}{2}rF^{l^4}\sigma^2_L\sigma^2_R\sigma^2_{\kappa}\right),
\end{equation}
with $u=1,\dots,F^{l+1}$. In Eq. \eqref{eq:init}, we can observe that the variance of $\Hb_{i,u}^{l+1,\mathrm{NL}}$ depends on the fourth power of the number of features. This term can easily become extremely large, therefore it is important to set $\sigma^2_L$, $\sigma^2_R$ and $\sigma^2_{\kappa}$ in such a way that they can balance it. In this work, we set $\sigma^2_L=\sigma^2_R=\frac{1}{F^{l^2}}$ and $\sigma^2_{\kappa}=\frac{2}{r}$. This allows us to obtain $\Hb_{i,u}^{l+1,\mathrm{NL}}\sim \Nc(0,1)$ with $u=1,\dots, F^{l+1}$.
\subsection{Analogy with unrolled graph smoothness optimization} \label{sec:analogy}

\begin{figure}
    \centering
    \subfigure[Linear inverse problem]{
    \includegraphics[width=0.232\textwidth]{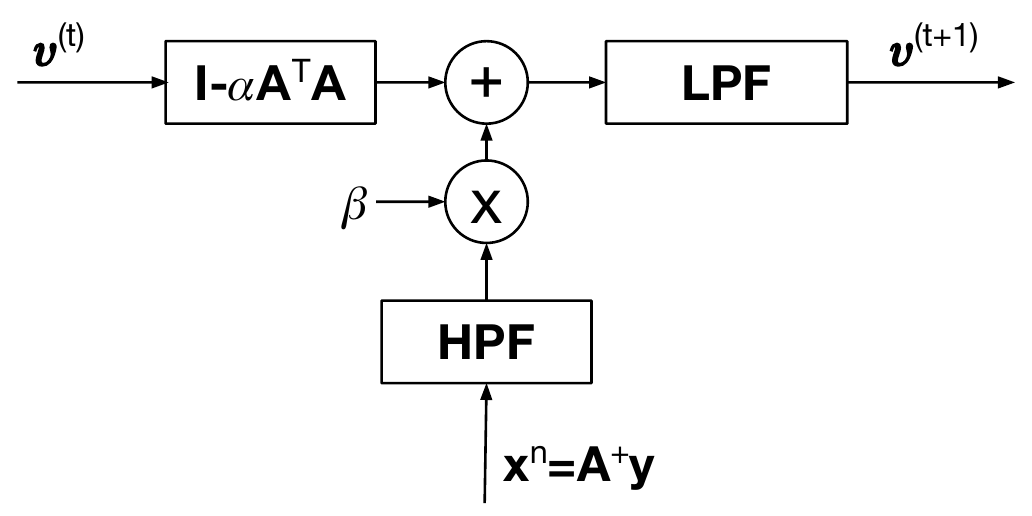}
    }
    \subfigure[Denoising]{
    \includegraphics[width=0.2\textwidth]{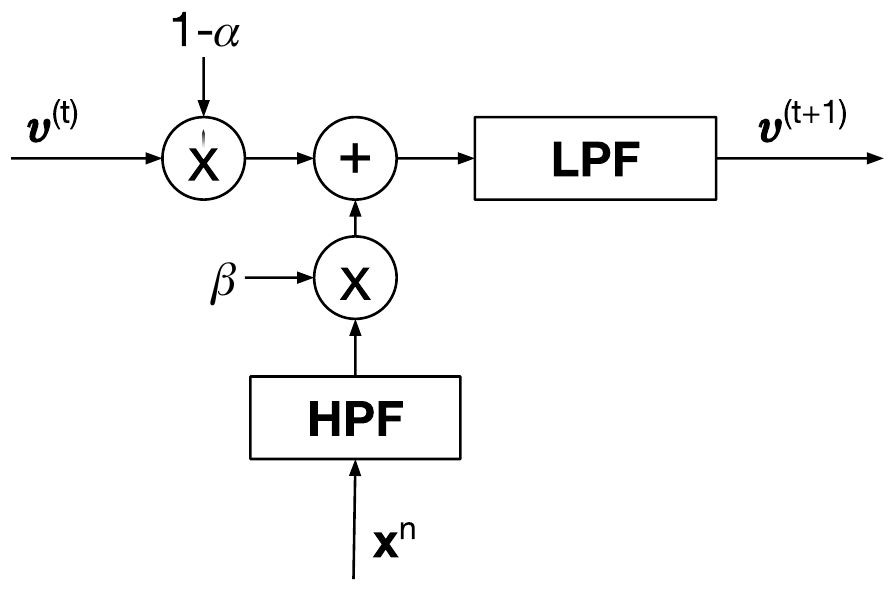}
    }
    \caption{Single iteration. LPF is graph lowpass filter, HPF is a graph highpass filter.}
    \label{fig:update_fig}
\end{figure}

The neural network architecture presented in Sec. \ref{sec:overview} can be seen as a generalization of few iterations of an unrolled proximal gradient descent optimization method, which is widely used to solve linear inverse problems in the form of
\begin{align}\label{eq:noise_model}
    \yb = \Ab\xb + \nb
\end{align}
being $\xb$ the clean image, $\Ab$ a forward model (e.g., a degradation such as blurring, downsampling, compressed sensing, etc.) and $\nb$ a noise term. A well-known technique to recover $\xb$ from $\yb$ is to cast the problem as a least-squares minimization problem with a regularization term that models some prior knowledge about the image. One such regularizer is graph smoothness. Considering a graph with Laplacian matrix $\Lb$ where edges connect pixels that are deemed correlated according to some criterion, the graph smoothness $\xb^T \Lb \xb$ is the graph equivalent of the total variation measure, indicating how much $\xb$ varies across the edges of the graph. Natural images where the graph connects the local neighborhood typically have lowpass behavior, resulting in a low graph smoothness value. Reconstruction is therefore cast as:
\begin{align} \label{eq:x_est}
    \hat{\xb} = \argmin{\xb} \left[ \frac{1}{2}\Vert \yb - \Ab\xb \Vert^2_2 + \frac{\beta}{2} \xb^T \Lb \xb \right]
\end{align}
The functional in Eq. \eqref{eq:x_est} is in the form of a sum of two terms ($f(\xb) + g(\xb)$) and can be minimized by means of proximal gradient descent \cite{combettes2011proximal} which alternates a gradient descent step over $f$ and a proximal mapping over $g$:
\begin{align*}
    \xb^{(t+1)} &= \text{prox}_g \left( \xb^{(t)} - \alpha \nabla_{\nub}f  \right) \\ &= \text{prox}_g \left( (\Ib-\alpha\Ab^T\Ab)\xb^{(t)} + \alpha \Ab^T\yb \right) \\
    \text{prox}_g \left( \mub \right) &= \argmin{\zb} \left[ \Vert \zb - \mub \Vert^2_2 + \frac{\beta}{2} \zb^T \Lb \zb \right].
\end{align*}
Solving for the proximal mapping operator results in the following update equation:
\begin{align}\label{eq:update}
    \xb^{(t+1)} =  \left( \Ib +\beta \Lb \right)^{-1}  \left[ (\Ib-\alpha\Ab^T\Ab)\xb^{(t)} + \alpha \Ab^T \yb \right].
\end{align}
In order to match the framework of residual networks, let us define the least-squares solution $\xb^n=\Ab^{+}\yb=\left( \Ab^T\Ab \right)^{-1}\Ab^T \yb$ and perform a change of variable whereby the optimization estimates the residual of the least squares solution, i.e., $\nub^{(t)}=\xb^n - \xb^{(t)}$. Hence, we can rewrite Eq. \eqref{eq:update} as:
\begin{align*}
    &\xb^n - \nub^{(t+1)} =\\&\left( \Ib +\beta \Lb \right)^{-1}  \left[ \left( \Ib-\alpha\Ab^T\Ab \right) \left( \xb^n-\nub^{(t)} \right) + \alpha \Ab^T \yb \right].
\end{align*}
Finally, the following update equation can be derived:
\begin{align}\label{eq:noise_est}
   \nub^{(t+1)} = \left( \Ib +\beta \Lb \right)^{-1}  \left[ \left( \Ib-\alpha\Ab^T\Ab \right) \nub^{(t)}  + \beta \Lb \xb^n \right].
\end{align}

This update can be visualized as in Fig. \ref{fig:update_fig}a and is composed of two major operations involving the signal prior:
\begin{enumerate}
    \item $\Lb \xb^n$: the graph Laplacian can be seen as a graph highpass filter applied to $\xb^n$;
    \item $\left( \Ib +\beta \Lb \right)^{-1}$: this term can be seen as a graph lowpass filter. In order to see this, let us use the matrix inversion lemma as $\left( \Ib +\beta \Lb \right)^{-1} = \left( \Ib +\beta \Ub \Lambdab \Ub^H \right)^{-1} = \Ib - \Ub \left( \beta^{-1}\Lambdab^{-1} + \Ib \right)^{-1} \Ub^H = \Ub \left[ \Ib - \left(\beta^{-1}\Lambdab^{-1} + \Ib \right)^{-1} \right] \Ub^H$, where $\Ub$ is the graph Fourier transform. The term $\Ib - \left(\beta^{-1}\Lambdab^{-1} + \Ib \right)^{-1}$ is a diagonal matrix whose entries are equal to $\frac{1}{\beta\lambda_i+1}$ where $\lambda_i$ are the eigenvalues of the graph Laplacian, and the lowpass behavior is due to decreasing value of such entries for increasing $\lambda$.
\end{enumerate}
For the denoising problem, we can set $\Ab=\Ib$ and obtain the update shown in Fig.\ref{fig:update_fig}b.
The network architecture proposed in Sec. \ref{sec:overview} draws from this derivation by unrolling a finite number of Eq. \eqref{eq:noise_est} iterations and generalizing the lowpass and highpass filters with learned graph filters interleaved by nonlinearities. In Sec. \ref{sec:experiments} we experimentally show that the learned filters actually show an approximate highpass and lowpass behavior.

\section{Experimental Results} \label{sec:experiments}

\subsection{Training details}
The training protocol follows the one used in \cite{zhang2017beyond}. The network is trained with patches of size $42 \times 42$ randomly extracted from 400 images from the train and test partitions of the Berkeley Segmentation Dataset (BSD) \cite{MartinFTM01}, withholding the 68 images in the validation set for testing purposes (BSD68 dataset). The loss function is the mean squared error (MSE) between the denoised patch output by the network and the ground truth. Each model is trained for approximately 800000 iterations with a batch size of 8. The Adam optimizer \cite{kingma2014adam} has been used with an exponentially decaying learning rate between $10^{-4}$ and $10^{-5}$. The behavior of the graph-convolutional layer is slightly different between training and testing for efficiency reasons. During training all pairwise distances are computed among the feature vectors corresponding to the pixels in the patch. On the other hand, testing is ``fully convolutional'', as every pixel has a search window centered around it and neighbors are identified as the closest pixels in such search window. The search window size is $43 \times 43$, roughly comparable to the patch size used in training. This procedure is slightly suboptimal as some pixels might suffer from border effects during training (their search windows are not centered around them) but it is advantageous in terms of speed and memory requirements. Reflection padding is used for all 2D convolutions to avoid border effects. The $\delta$ parameter in the edge attention term in Eq. \eqref{eq:edge_attention} is set to a value equal to 10. The number of features used in all convolutional layers is 132, except for the three parallel branches of the preprocessing stage which have 44 features. The number of
circulant rows in the circulant approximation of dense layers in the $\fun$ network is $m=3$. The low-rank approximation uses $r=11$ terms. During training, we noticed that the proposed lightweight ECC presented in Sec. \ref{sec:ecc} is extremely useful. In fact, without it, the network suffered from vanishing gradient problems even with a significantly lower number of layers.

\subsection{Feature analysis}
In this section we study the properties of the features in the hidden layers of the network.

\subsubsection{Adaptive receptive field}
\begin{figure}
\centering
\includegraphics[width=0.32\columnwidth]{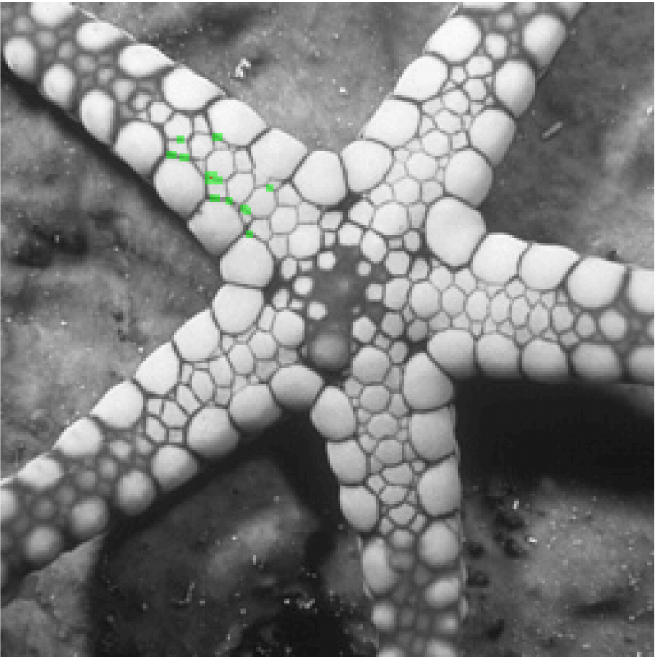}
\includegraphics[width=0.32\columnwidth]{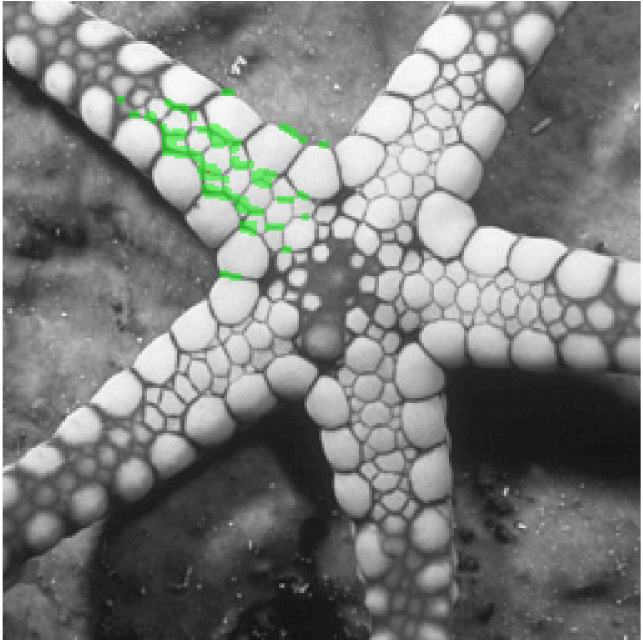}
\includegraphics[width=0.32\columnwidth]{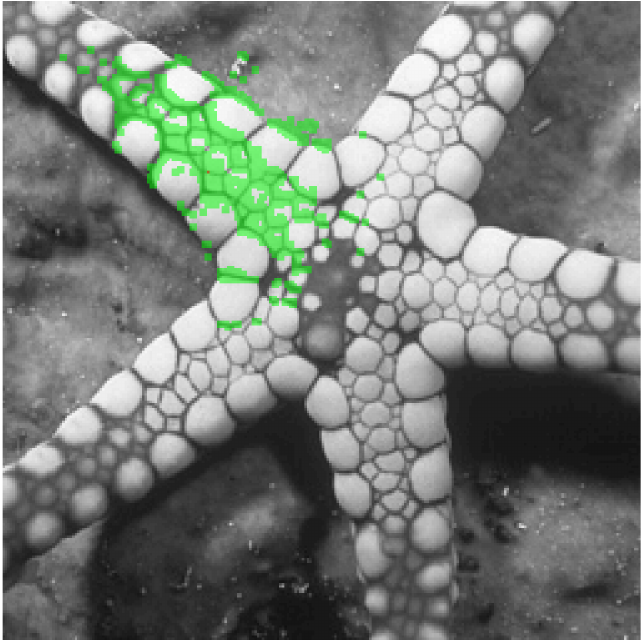}
\\[6pt]
\includegraphics[width=0.32\columnwidth]{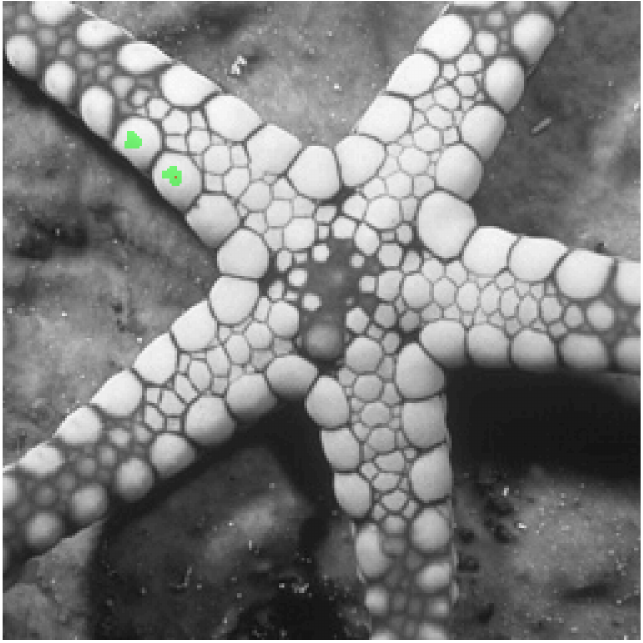}
\includegraphics[width=0.32\columnwidth]{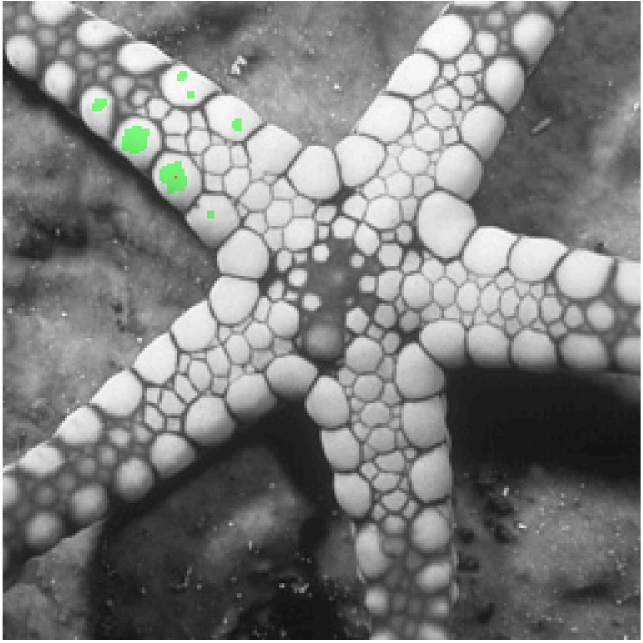}
\includegraphics[width=0.32\columnwidth]{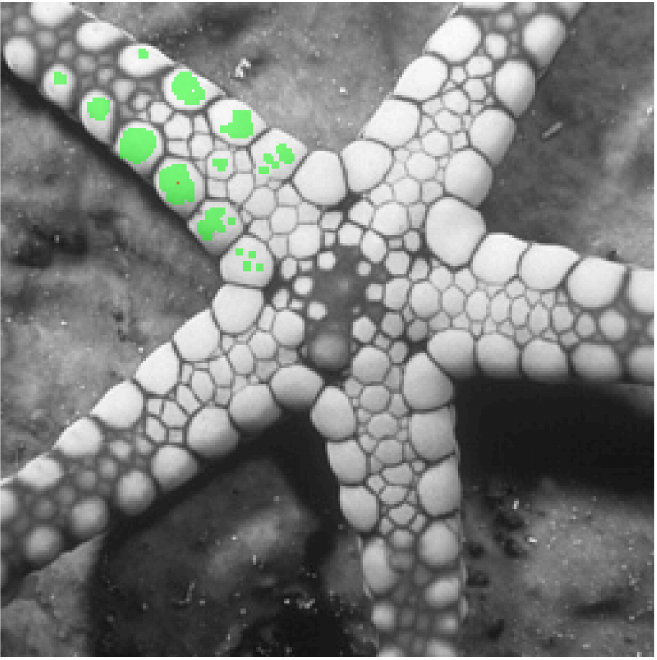}
\caption{Receptive field (green) of a single pixel (red) for the three graph-convolutional layers in the LPF$_1$ block with respect to the input of the first graph-convolutional layer in the block. Top row: gray pixel on an edge. Bottom row: white pixel in a uniform area.}
\label{fig:recfield}
\end{figure}

\begin{figure*}
\centering
\includegraphics[width=0.119\textwidth]{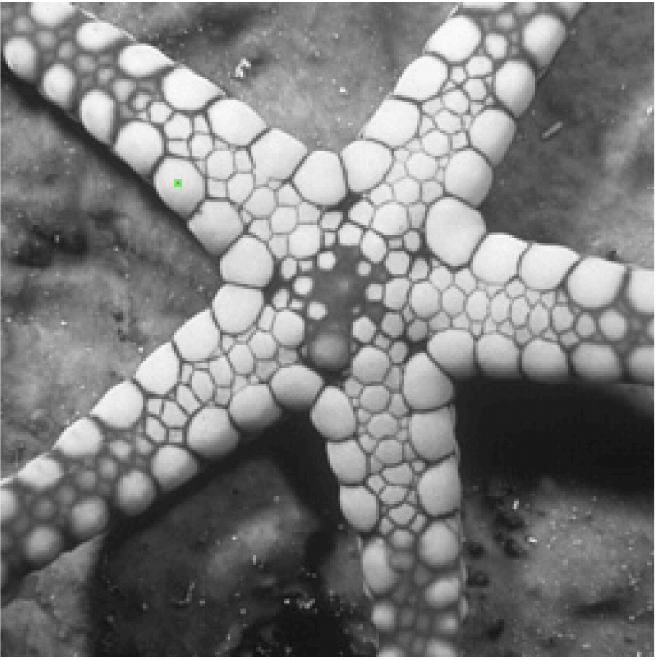}
\includegraphics[width=0.119\textwidth]{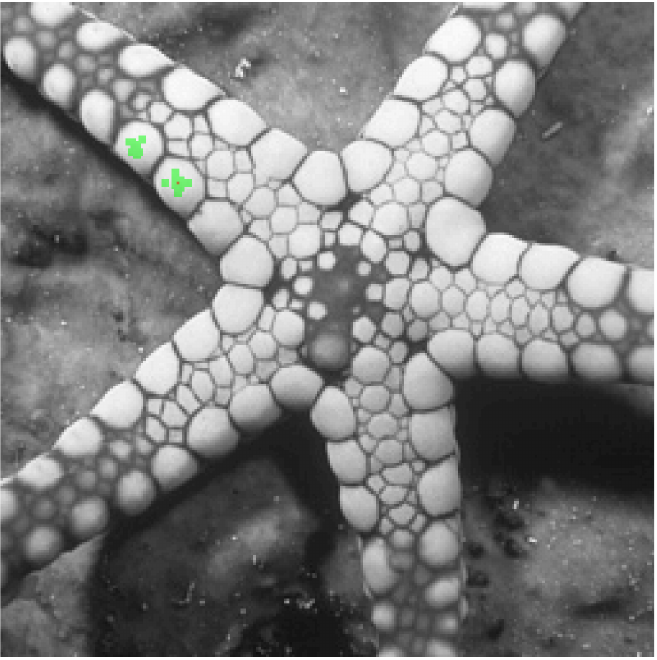}
\includegraphics[width=0.119\textwidth]{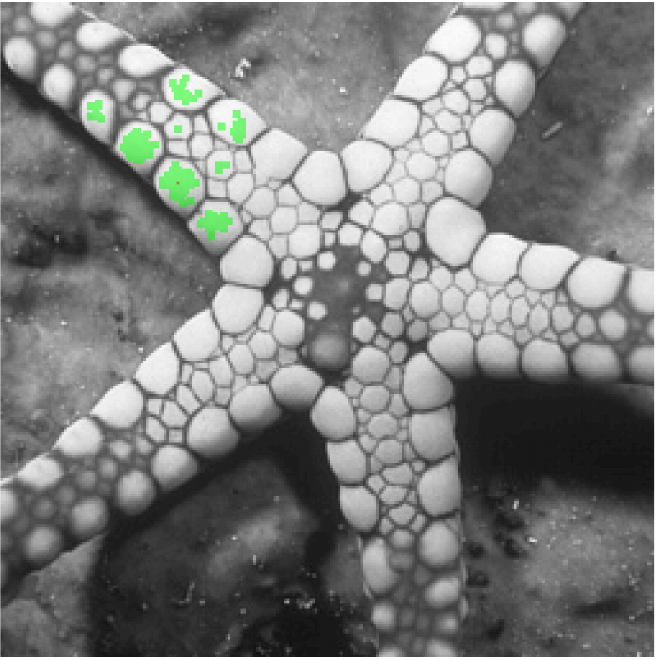}
\includegraphics[width=0.119\textwidth]{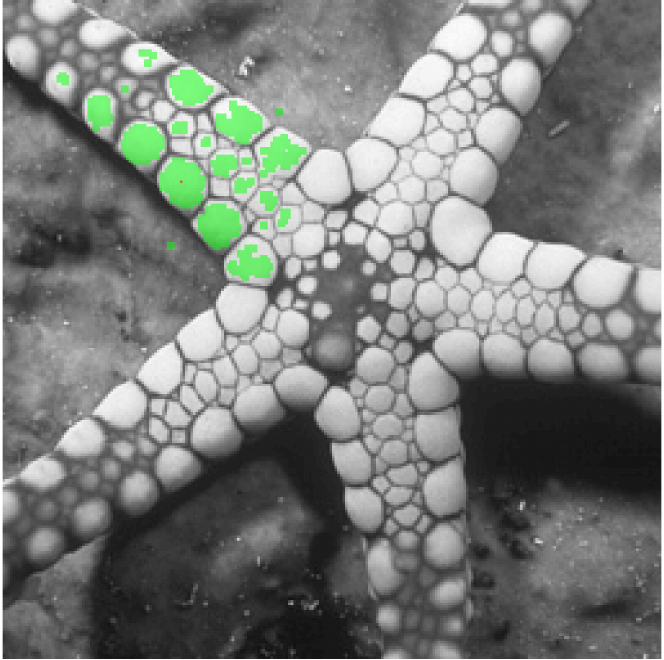}
\includegraphics[width=0.119\textwidth]{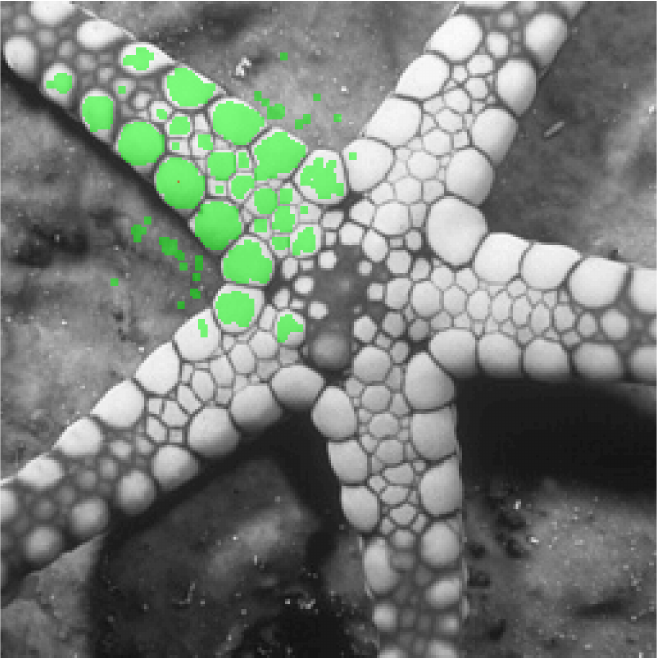}
\includegraphics[width=0.119\textwidth]{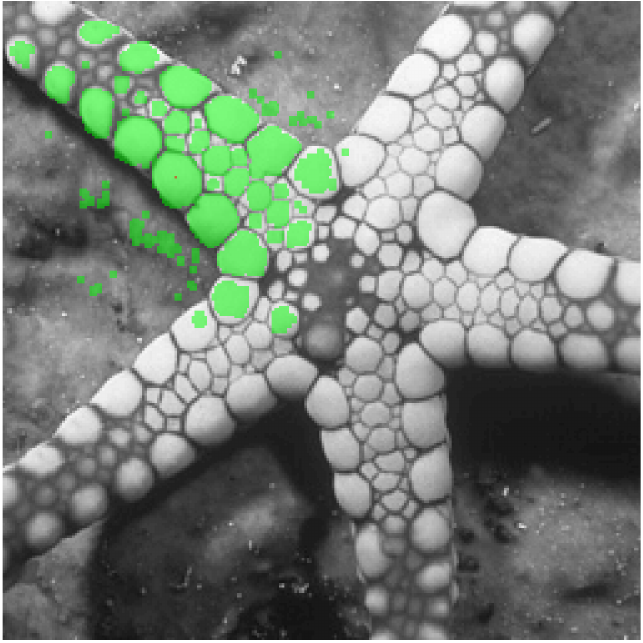}
\includegraphics[width=0.119\textwidth]{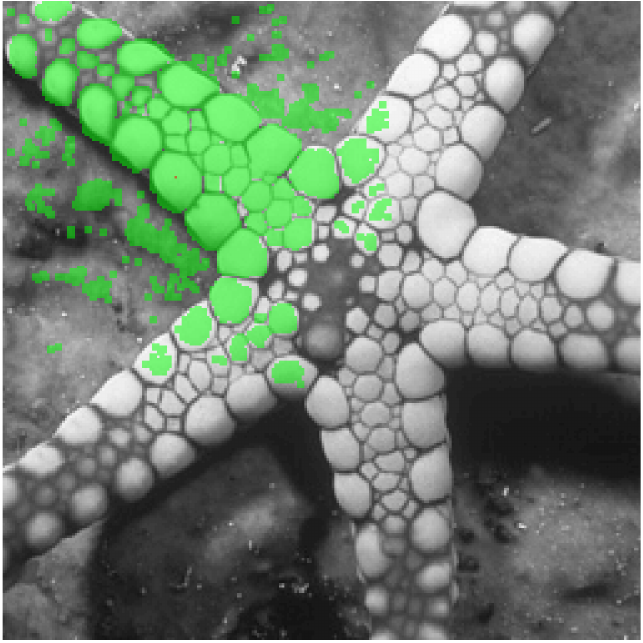}
\includegraphics[width=0.119\textwidth]{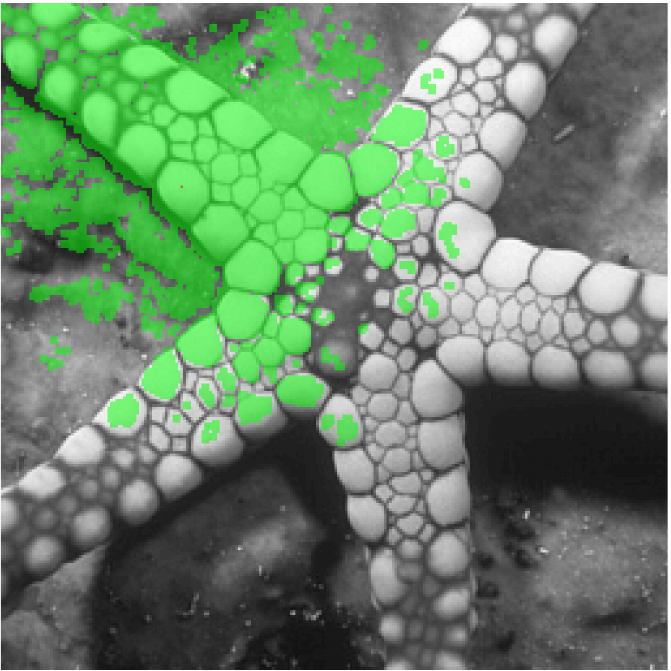}
\caption{Receptive field (green) of a single pixel (red) for the layers in the HPF and LPF$_1$ blocks in the same order as a forward pass, with respect to the output of the preprocessing block.}
\label{fig:recfield_long}
\end{figure*}

\begin{figure}
\centering
\includegraphics[width=0.3\columnwidth]{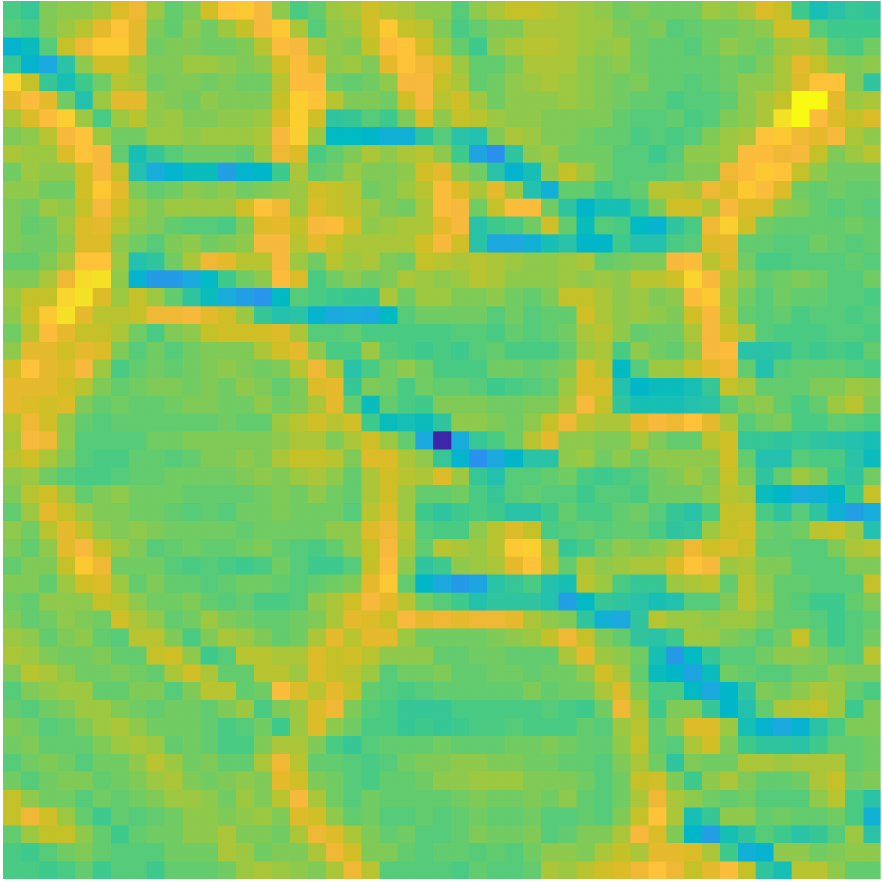}
\includegraphics[width=0.3\columnwidth]{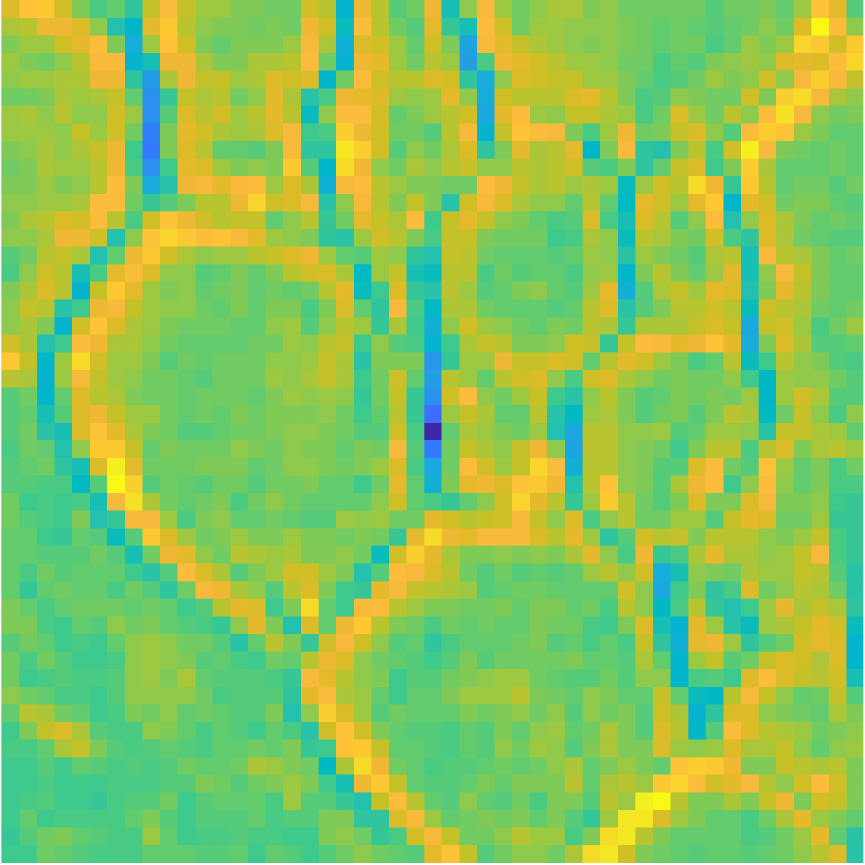}
\includegraphics[width=0.3\columnwidth]{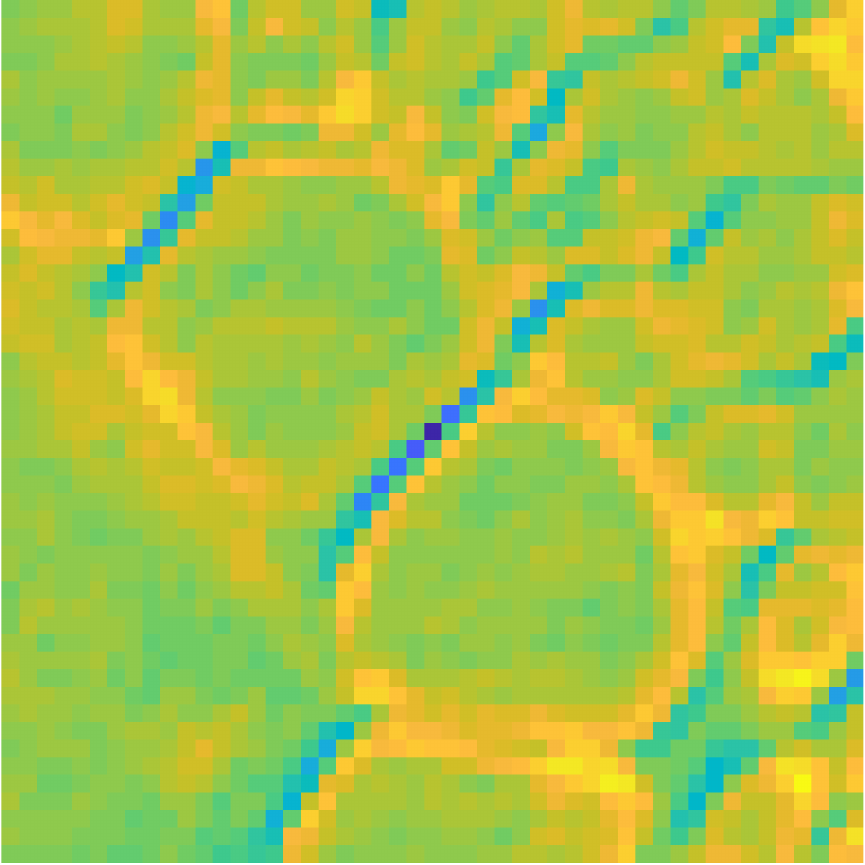}
\caption{Euclidean distances between feature vectors of the central pixel and all the pixels in the search window (input of first graph-convolutional layer of LPF$_1$). Left to right: pixel on a horizontal edge, pixel on a vertical edge, pixel on a diagonal edge. Blue represents lower distance.}
\label{fig:dir}
\end{figure}

We first analyze the characteristics of the receptive field of a single pixel. Since the proposed network employs graph-convolutional layers, the shape of the receptive field is not fixed as in classical CNNs, but it depends on the structure of the graph. In Fig. \ref{fig:recfield} we show two examples of the receptive field of a single pixel for the graph-convolutional layers in an LPF block with respect to the input of the block. Instead, in Fig. \ref{fig:recfield_long} we show the receptive field of a single pixel for the layers in the HPF and in the first LPF blocks with respect to the output of the preprocessing block. We can clearly see that the receptive field is adapted to the characteristics of the image: if we consider a pixel in a uniform area, its receptive field will mostly contain pixels that belong to similar regions; instead if we consider a pixel on an edge, its receptive field will be mainly composed of other edge pixels. This is beneficial to the denoising task as it allows to exploit self-similarity and it descends from the use of a nearest neighbor graph, connecting each pixel to other pixels with similar features. Notice that differently from algorithms performing block matching in the pixel space, we compute distances between feature vectors which can capture more complex image characteristics. This can be seen in Fig. \ref{fig:dir} where we compute the Euclidean distances between the feature vector of the central pixel and the feature vectors of the other pixels in the search window. We notice that the distances reflect the type of edge that includes the central pixel, e.g., a pixel sitting on a horizontal edge will detect as closest other pixels sitting on horizontal edges. This is due to the visual features learned by the network and would not happen in pixel-space matching.
Thanks to the adaptability of the receptive field, graph convolution can be interpreted as a generalization of the block matching operation performed in other non-local denoising methods, such as BM3D \cite{dabov2007image}.


\begin{figure}
\centering
\includegraphics[width=0.32\columnwidth]{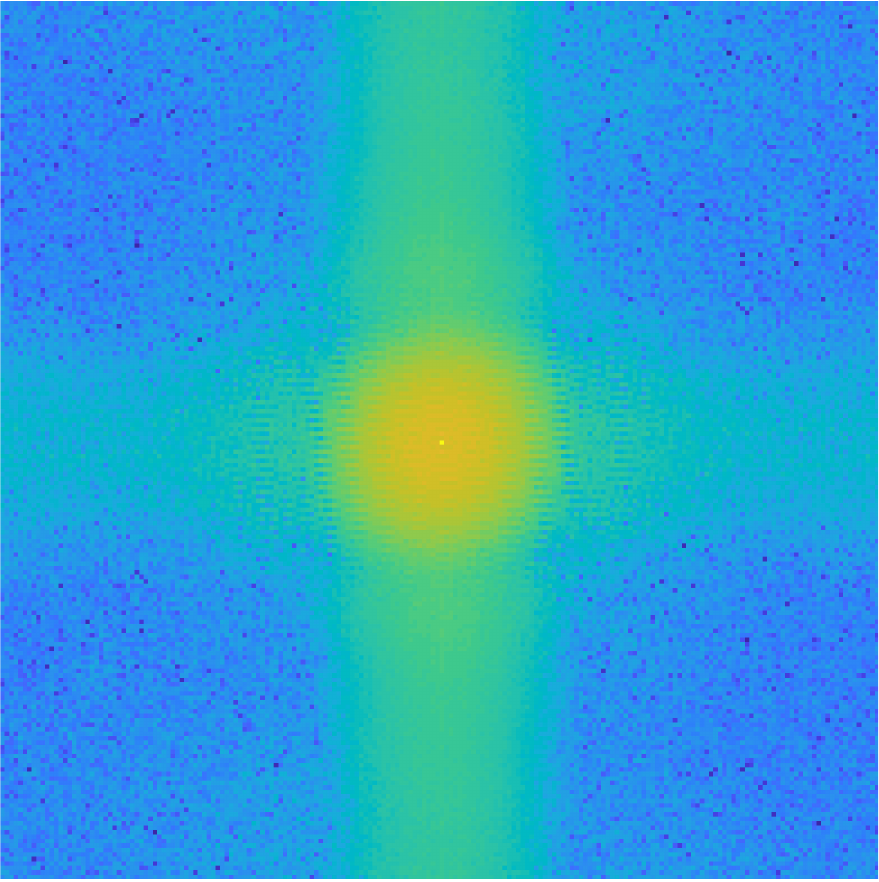}
\includegraphics[width=0.32\columnwidth]{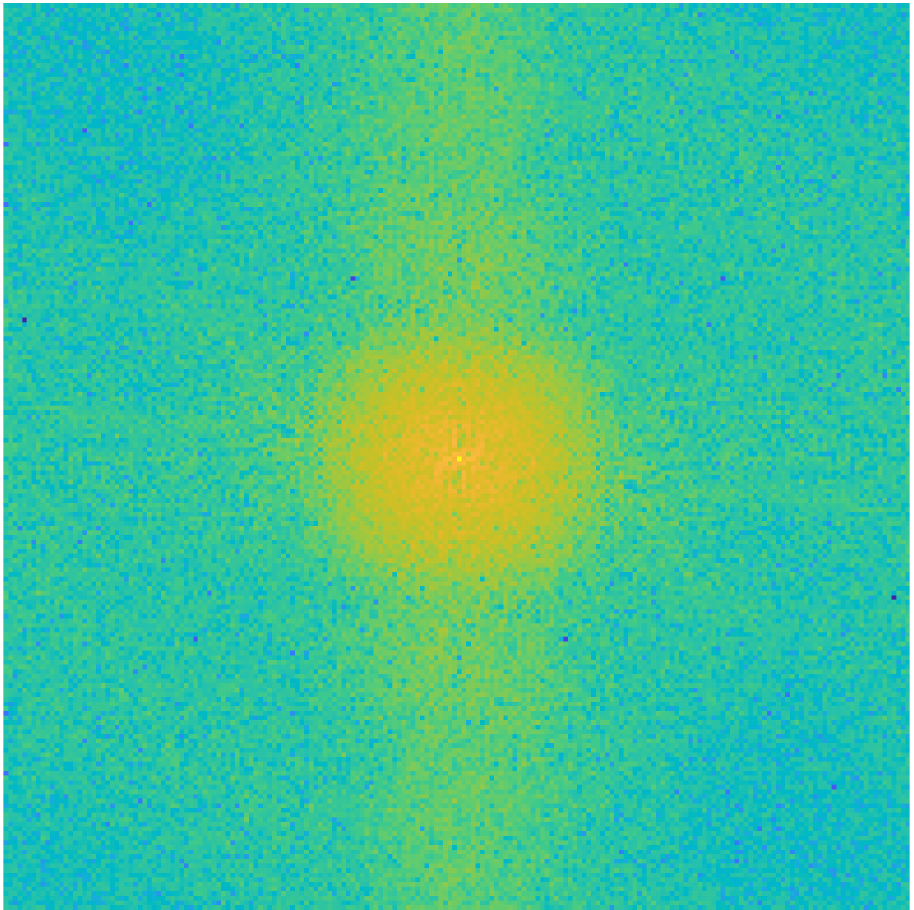}
\includegraphics[width=0.32\columnwidth]{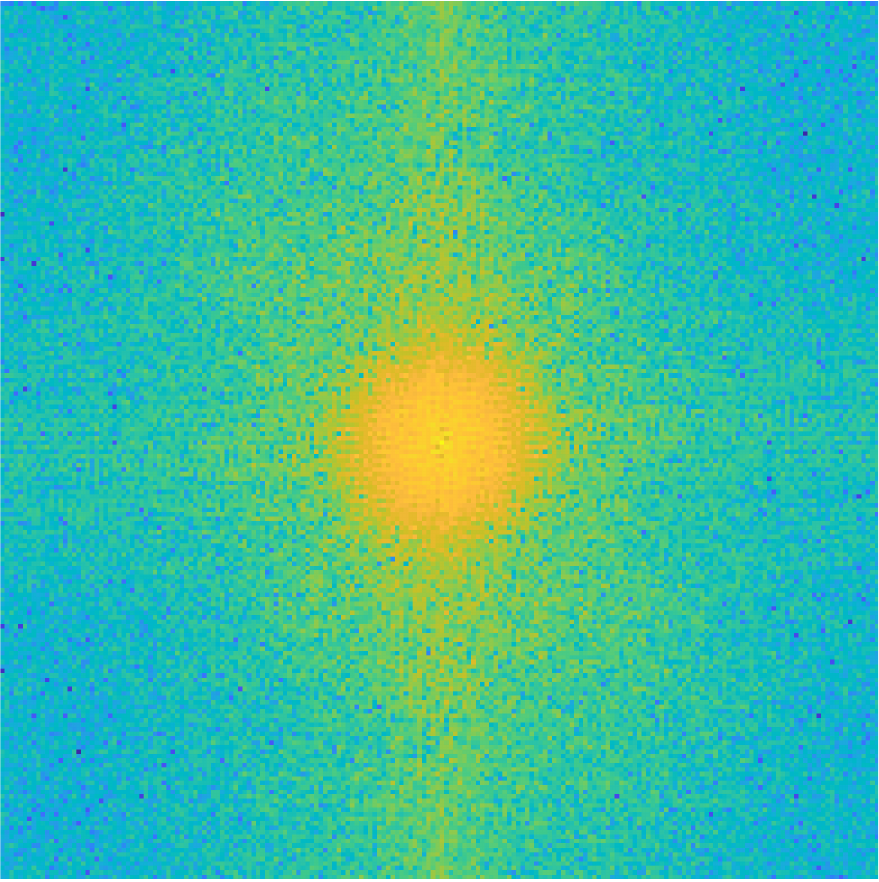}
\\[6pt]
\includegraphics[width=0.32\columnwidth]{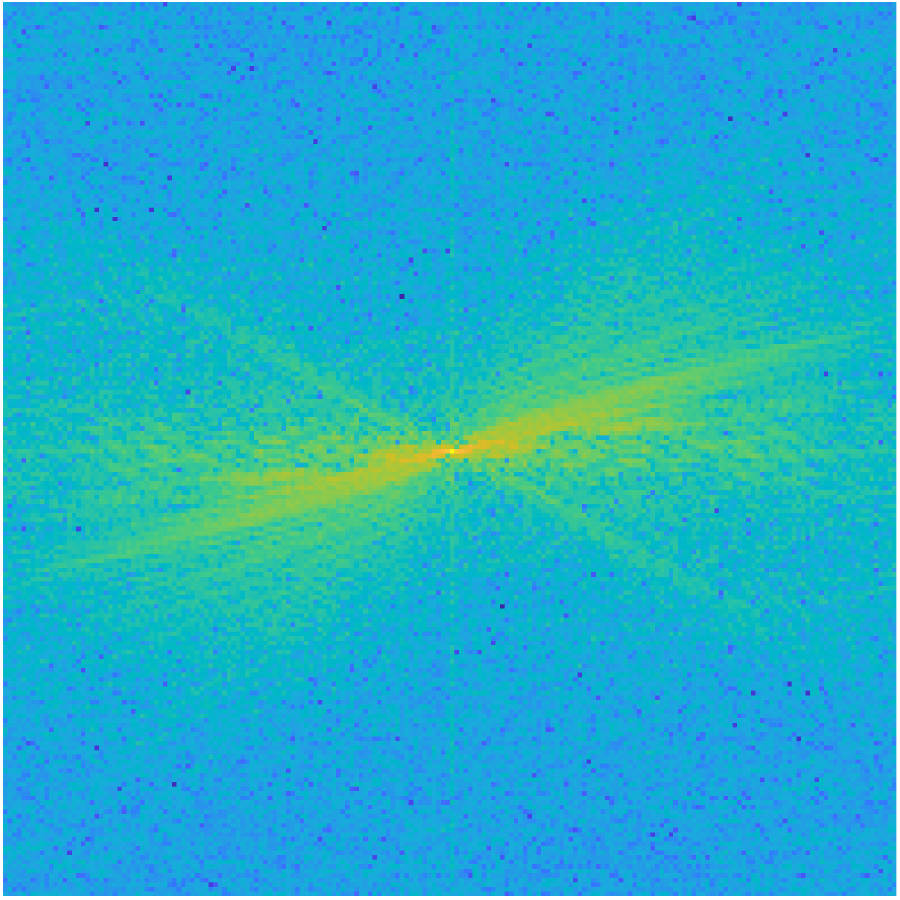}
\includegraphics[width=0.32\columnwidth]{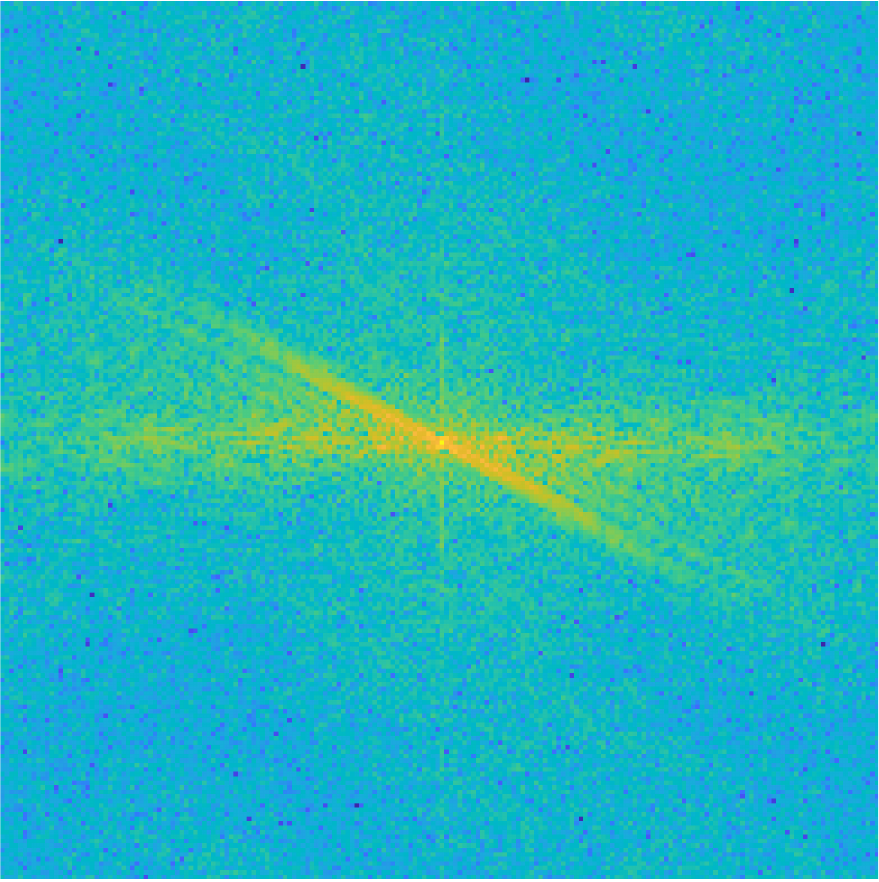}
\includegraphics[width=0.32\columnwidth]{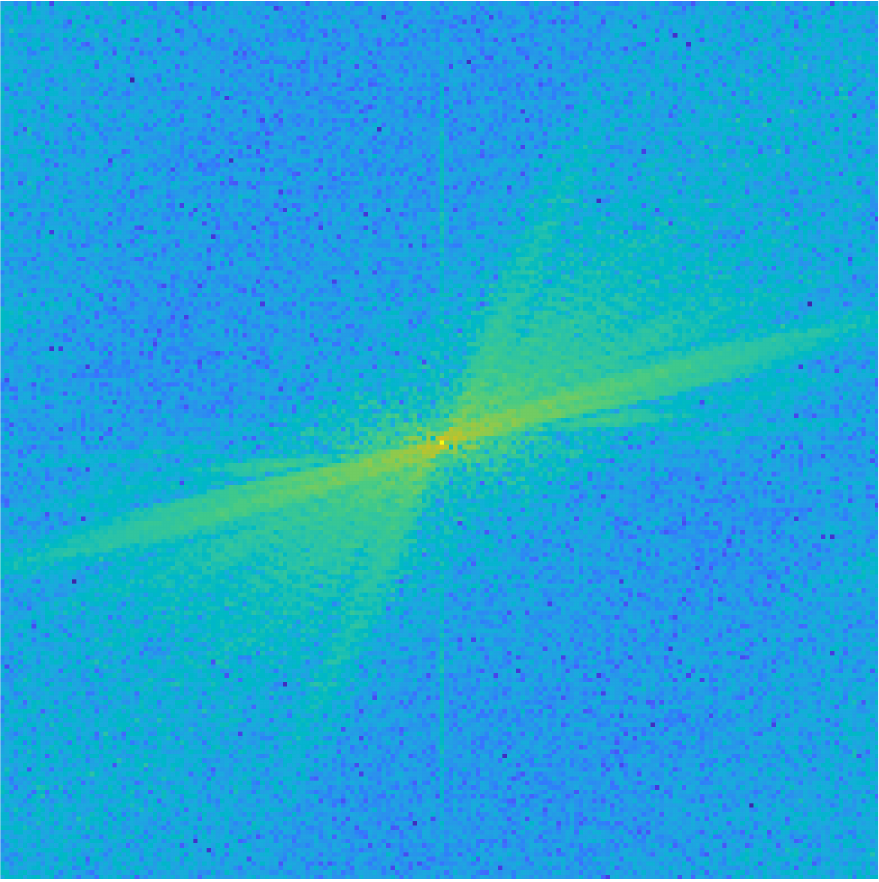}
\caption{Log-magnitude of discrete Fourier transform of three feature maps at the output of the LPF$_1$ block (top) and HPF block (bottom). Blue is lower magnitude.}
\label{fig:filters}
\end{figure}

\begin{figure}
\centering
\includegraphics[width=0.9\columnwidth]{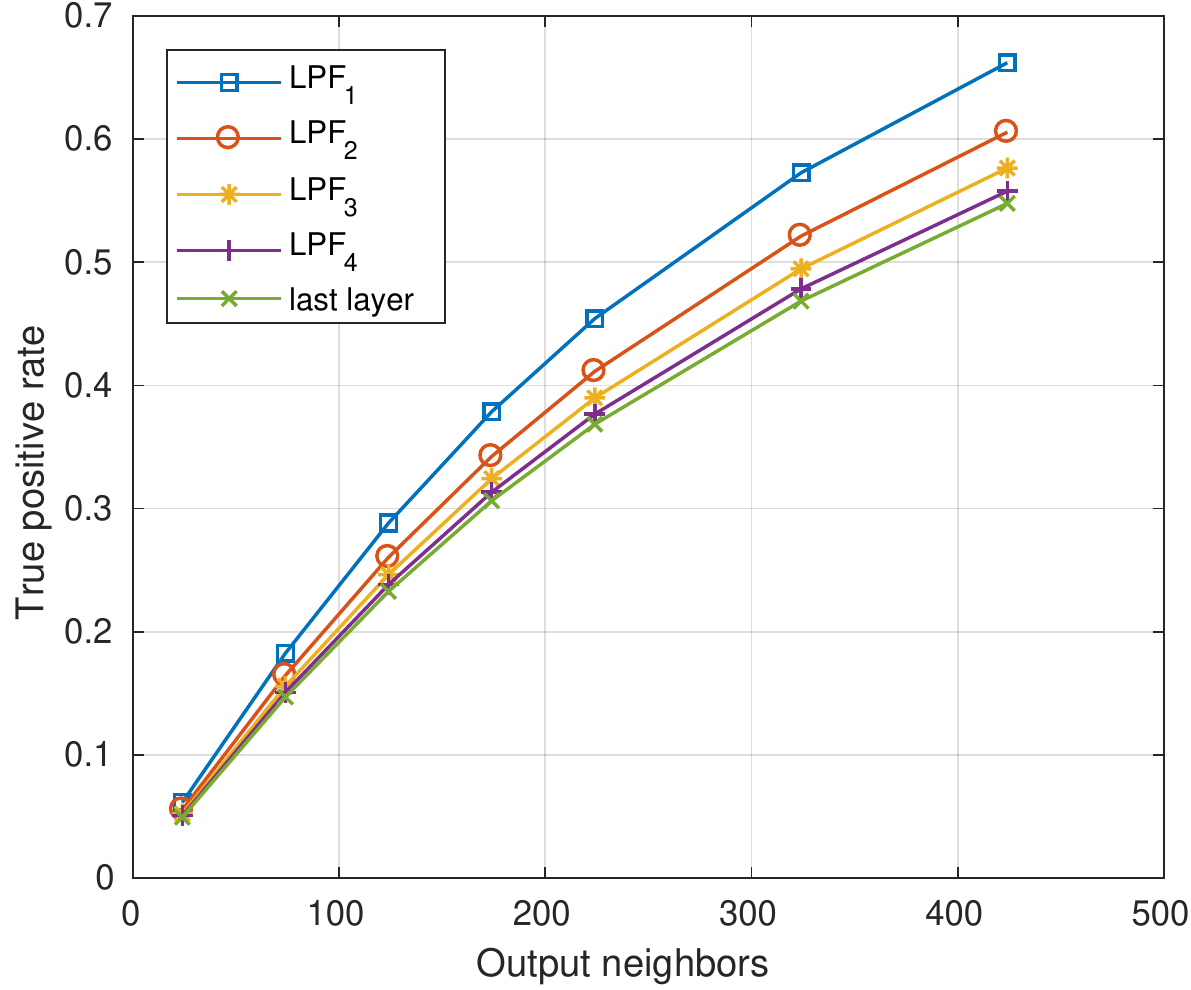}
\caption{Accuracy of edge prediction from hidden layers.}
\label{fig:graph_embedding}
\end{figure}

\subsubsection{Filter analysis}
We also study the behavior of the LPF and HPF operators. In particular, we are interested in validating the analogy made in Sec. \ref{sec:analogy}. We compute the discrete Fourier transform (DFT) of the feature maps at the output of these operators. As an example, Fig. \ref{fig:filters} shows the log-magnitude of the coefficients of three feature maps at the output of the HPF block and of the first LPF block. The energy of the DFT coefficients of the LPF feature maps is concentrated in the low frequencies, thus showing a lowpass behavior. Instead, the coefficients of the HPF feature maps show a typical highpass behavior, having the energy concentrated along few directions.
This substantiates our claim that the learned convolutional layers actually approximate nonlinear highpass and lowpass operators.

\subsubsection{Edge prediction}
Lastly, we measure how much the true graph constructed by pixel or patch similarities on the noiseless image is successfully predicted by the graph constructed from the feature vectors in the hidden layers. In order to construct the true graph of the image, we first compute the average pixel value of a 5$\times$5 window centered at the considered pixel, for every pixel in the image, and then we use the obtained values to compute a nearest neighbor graph with Euclidean distances. We then compare the true graph with the graph computed in the hidden layers of the network. Fig. \ref{fig:graph_embedding} shows the percentage of edges correctly identified as function of the number of neighbors considered for the true graph. We can notice that the accuracy of the prediction decreases in layers closer to the output. This is due to the fact that we use a residual network that estimates the noise instead of approximating the clean image. In fact, the network learns to successively remove the latent correlations in the feature space, and as a consequence, the graph becomes more random in the later layers.

\subsection{Ablation studies}
We study the impact of various design parameters on denoising performance. First, Table \ref{table:psnr_vs_nn} shows the PSNR on the Set12 testing set as function of the number of neighbors used by the graph convolution operation for several values of the noise standard deviation $\sigma$. Each model has been independently trained for the specified number of neighbors.  It can be noticed that increasing the number of neighbors improves the denoising performance up to a saturation point, and then the performance slightly decreases. This shows that there an optimal neighborhood size and that it is important to employ only a small number of neighbors, in order to select only pixels with similar characteristics. This is in contrast with the NLRN method which uses all the pixels in the search window.

Then, we study the relative impact on performance of the edge aggregation matrices $\bm{\Theta}$ in Eq. \eqref{eq:ecc} with respect to using only the edge attention scalar $\gamma$. Table \ref{table:onlyscalar} reports the PSNR achieved on Set12 by the proposed method with the non-local aggregation performed as in Eq. \eqref{eq:ecc} and a variant where the aggregation is computed as:
\begin{align*}
    \Hb_i^{l+1,\mathrm{NL}}  &=  \sum_{j\in\Sx_i^l} \gamma^{j\to i} \Hb_j^l .
\end{align*}
Both methods use a non-local graph with 8 nearest neighbors.
We can notice that the edge attention term alone achieves a worse PSNR with respect to GCDN by approximately 0.2 dB, even though it improves over a model without non-local neighbors (see Table \ref{table:psnr_vs_nn} for the corresponding 0-NN value). This shows the advantage of using a trainable affine transformation, such as $\bm{\Theta}$ in Eq. \eqref{eq:ecc}, instead of a scalar weight function with a predefined structure. 

Finally, we remark that we do not compare with respect to the full ECC without the approximations introduced in Sec. \ref{sec:ecc} because it suffers from vanishing gradient problems, rendering training unstable even for a much smaller number of layers, and it would be computationally prohibitive.

\begin{table} 
\centering
\caption{PSNR (dB) v. non-local neighborhood size (Set12)}
\begin{tabular}{ccccccc}
\hline
$\sigma$ & 0-NN & 4-NN & 8-NN & 12-NN & 16-NN & 20-NN \\ \hline
15       & 32.91 & 33.09 & 33.11 & 33.13 & 33.14 & 33.13 \\ \hline
25       & 30.50 & 30.70 & 30.74 & 30.75 & 30.78 & 30.78 \\ \hline
50       & 27.28 & 27.52 & 27.58 & 27.58 & 27.60 & 27.59 \\ \hline
\end{tabular}
\label{table:psnr_vs_nn}
\end{table}

\begin{table}
\centering
\caption{Edge attention v. ECC + edge attention (8-NN). PSNR (dB).}
\begin{tabular}{ccc}
\hline
$\sigma$ & Edge attention only & Proposed \\ \hline
25       & 30.53 & 30.74 \\ \hline
\end{tabular}
\label{table:onlyscalar}
\end{table}

\subsection{Comparison with state of the art}

\begin{table*}[t]
\centering
\caption{Natural image denoising results. Metrics are PNSR (dB) and SSIM.}
\begin{tabular}{ccccccccc}
\hline
Dataset                   & Noise $\sigma$ & BM3D           & WNNM           & TNRD           & DnCNN          & N$^3$Net  & NLRN                    & \textbf{GCDN}       \\ \hline
\multirow{3}{*}{Set12}    & 15             & 32.37 / 0.8952 & 32.70 / 0.8982 & 32.50 / 0.8958 & 32.86 / 0.9031 & - / -     & \textbf{33.16} / 0.9070 & 33.14 / \textbf{0.9072}         \\
                          & 25             & 29.97 / 0.8504 & 30.28 / 0.8557 & 30.06 / 0.8512 & 30.44 / 0.8622 & 30.55 / - & \textbf{30.80 / 0.8689} & 30.78 / 0.8687          \\
                          & 50             & 26.72 / 0.7676 & 27.05 / 0.7775 & 26.81 / 0.7680 & 27.18 / 0.7829 & 27.43 / - & \textbf{27.64 / 0.7980} & 27.60 / 0.7957          \\ \hline
\multirow{3}{*}{BSD68}    & 15             & 31.07 / 0.8717 & 31.37 / 0.8766 & 31.42 / 0.8769 & 31.73 / 0.8907 & - / -     & \textbf{31.88} / 0.8932 & 31.83 / \textbf{0.8933}          \\ 
                          & 25             & 28.57 / 0.8013 & 28.83 / 0.8087 & 28.92 / 0.8093 & 29.23 / 0.8278 & 29.30 / - & \textbf{29.41} / 0.8331 & 29.35 / \textbf{0.8332} \\ 
                          & 50             & 25.62 / 0.6864 & 25.87 / 0.6982 & 25.97 / 0.6994 & 26.23 / 0.7189 & 26.39 / - & \textbf{26.47} / 0.7298 & 26.38 / \textbf{0.7389}          \\ \hline
\multirow{3}{*}{Urban100} & 15             & 32.35 / 0.9220 & 32.97 / 0.9271 & 31.86 / 0.9031 & 32.68 / 0.9255 & - / -     & 33.42 / 0.9348          & \textbf{33.47 / 0.9358} \\ 
                          & 25             & 29.70 / 0.8777 & 30.39 / 0.8885 & 29.25 / 0.8473 & 29.97 / 0.8797 & 30.19 / - & 30.88 / 0.9003 & \textbf{30.95 / 0.9020}         \\ 
                          & 50             & 25.95 / 0.7791 & 26.83 / 0.8047 & 25.88 / 0.7563 & 26.28 / 0.7874 & 26.82 / - & 27.40 / \textbf{0.8244} & \textbf{27.41} / 0.8160          \\ \hline
\end{tabular}
\label{table:results}
\end{table*}

\begin{table*}[t]
\setlength{\tabcolsep}{4pt} 
\renewcommand{\arraystretch}{1} 
\centering
\caption{Depth map denoising results. Metrics are PNSR (dB) and SSIM.}
\begin{tabular}{ccccccccccc}
\hline
$\sigma$      & Method   & \textit{aloe}           & \textit{art}            & \textit{baby}           & \textit{cones}          & \textit{dolls}          & \textit{laundry}        & \textit{moebius}        & \textit{reindeer}               & Average                 \\ \hline
\multirow{3}{*}{15} & GCDN & 40.74 / \textbf{0.9873}          & 40.66 / \textbf{0.9886}          & 41.64 / \textbf{0.9917}          & 39.29 / \textbf{0.9832}          & \textbf{40.70 / 0.9830} & \textbf{41.97 / 0.9842} & \textbf{42.07 / 0.9877}   & \textbf{42.62 / 0.9915}           & \textbf{41.21 / 0.9872} \\ 
                    & NLRN     & 40.50 / 0.9844          & 40.48 / 0.9858          & \textbf{41.76} / 0.9899 & 39.50 / 0.9814          & 40.69 / 0.9800          & 41.96 / 0.9814          & 42.01 / 0.9848          & 42.44 / 0.9880           & 41.17 / 0.9845          \\
                    & OGLR     & \textbf{40.82} / 0.9801 & \textbf{40.77} / 0.9821 & 40.90 / 0.9806          & \textbf{39.65} / 0.9774 & 40.41 / 0.9756          & 41.32 / 0.9764          & 41.48 / 0.9793          & 41.72 / 0.9823                   & 40.88 / 0.9792          \\ \hline
\multirow{3}{*}{25} & GCDN & \textbf{37.12 / 0.9771} & \textbf{37.15 / 0.9788} & \textbf{37.50 / 0.9814} & 35.88 / \textbf{0.9697}          & \textbf{37.05 / 0.9705} & \textbf{38.62 / 0.9730} & \textbf{38.39 / 0.9786} & \textbf{38.80 / 0.9836}          & \textbf{37.56 / 0.9766} \\ 
                    & NLRN     & 37.08 / 0.9720          & 37.01 / 0.9734          & 37.37 / 0.9797          & \textbf{36.09} / 0.9661 & 37.01 / 0.9646          & 38.42 / 0.9679          & 38.33 / 0.9723          & 38.65 / 0.9786         & 37.50 / 0.9718          \\ 
                    & OGLR     & 36.67 / 0.9592          & 36.68 / 0.9649          & 36.29 / 0.9594          & 35.51 / 0.9545          & 36.41 / 0.9541          & 37.44 / 0.9541          & 37.17 / 0.9575          & 37.86 / 0.9655                & 36.75 / 0.9587          \\ \hline
\multirow{3}{*}{50} & GCDN & \textbf{33.37 / 0.9522} & \textbf{33.18 / 0.9536} & 32.23 / 0.9468          & \textbf{31.61 / 0.9379} & 32.37 / \textbf{0.9417}          & 34.07 / \textbf{0.9526}          & \textbf{33.73 / 0.9567} & 34.35 / \textbf{0.9672}                   & \textbf{33.11 / 0.9511} \\  
                    & NLRN     & 33.23 / 0.9444          & 32.86 / 0.9448          & \textbf{32.42 / 0.9534} & 31.53 / 0.9304          & \textbf{32.40} / 0.9347 & \textbf{34.15} / 0.9459 & 33.58 / 0.9475          & \textbf{34.37} / 0.9603  & 33.07 / 0.9452          \\ 
                    & OGLR     & 32.24 / 0.9121          & 31.92 / 0.9129          & 31.23 / 0.9027          & 30.21 / 0.8926          & 31.44 / 0.8999          & 32.85 / 0.9051          & 32.46 / 0.9093          & 32.99 / 0.9191               & 31.92 / 0.9067          \\ \hline
\end{tabular}
\label{table:depthmap}
\end{table*}

\begin{figure*}
\vspace*{5cm}
\centering
\parbox{0.161\textwidth}{%
\begin{tikzpicture}[overlay,remember picture,spy using outlines={rectangle,magnification=3.5,size=2cm}]
    \node[inner sep=0pt, anchor=south west,outer sep=0pt] 
    {\pgfimage[width=0.161\textwidth]{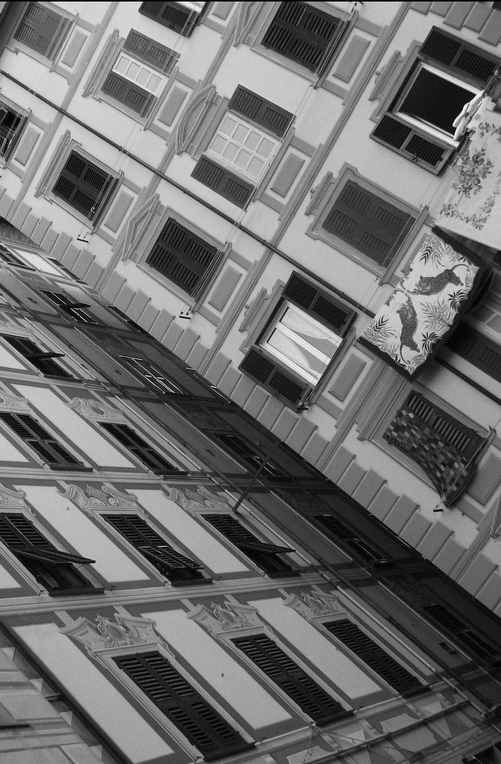}};
    {\spy[red!70!black] on (0.7,0.7) in node at (1.00,3.51);}
\end{tikzpicture}
}
\parbox{0.161\textwidth}{%
\begin{tikzpicture}[overlay,remember picture,spy using outlines={rectangle,magnification=3.5,size=2cm}]
    \node[inner sep=0pt, anchor=south west,outer sep=0pt] 
    {\pgfimage[width=0.161\textwidth]{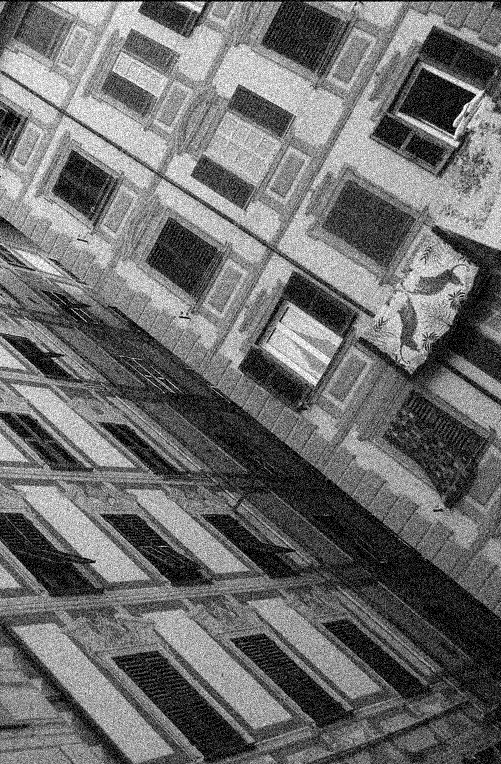}};
    {\spy[red!70!black] on (0.7,0.7) in node at (1.00,3.51);}
\end{tikzpicture}
}
\parbox{0.161\textwidth}{
\begin{tikzpicture}[overlay,remember picture,spy using outlines={rectangle,magnification=3.5,size=2cm}]
    \node[inner sep=0pt, anchor=south west,outer sep=0pt] 
    {\pgfimage[width=0.161\textwidth]{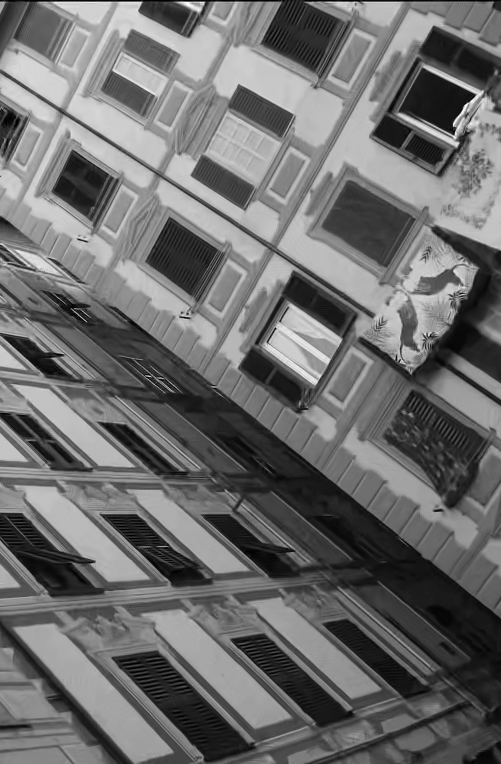}};
    {\spy[red!70!black] on (0.7,0.7) in node at (1.00,3.51);}
\end{tikzpicture}
}
\parbox{0.161\textwidth}{
\begin{tikzpicture}[overlay,remember picture,spy using outlines={rectangle,magnification=3.5,size=2cm}]
    \node[inner sep=0pt, anchor=south west,outer sep=0pt] 
    {\pgfimage[width=0.161\textwidth]{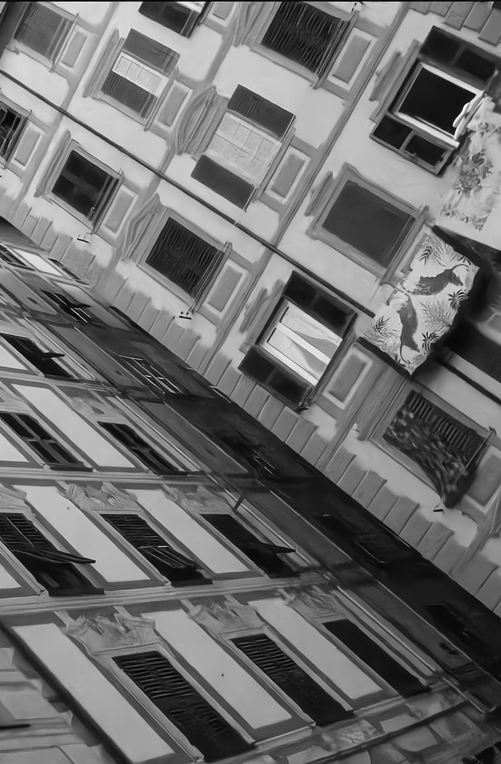}};
    {\spy[red!70!black] on (0.7,0.7) in node at (1.00,3.51);}
\end{tikzpicture}
}
\parbox{0.161\textwidth}{
\begin{tikzpicture}[overlay,remember picture,spy using outlines={rectangle,magnification=3.5,size=2cm}]
    \node[inner sep=0pt, anchor=south west,outer sep=0pt] 
    {\pgfimage[width=0.161\textwidth]{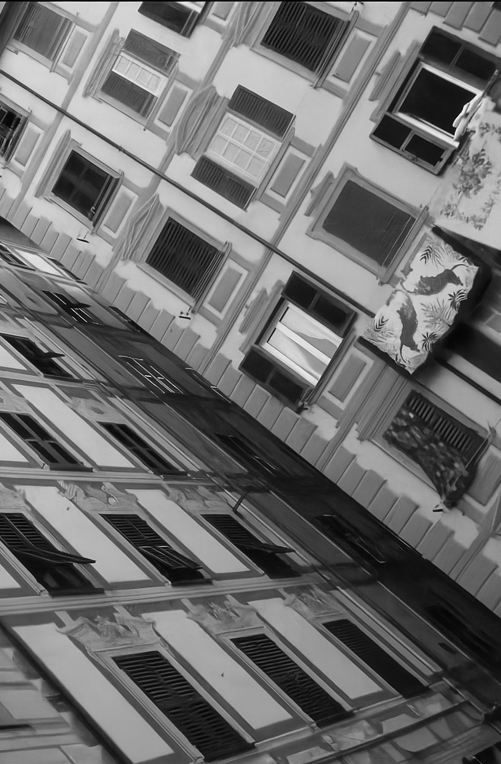}};
    {\spy[red!70!black] on (0.7,0.7) in node at (1.00,3.51);}
\end{tikzpicture}
}
\parbox{0.161\textwidth}{
\begin{tikzpicture}[overlay,remember picture,spy using outlines={rectangle,magnification=3.5,size=2cm}]
    \node[inner sep=0pt, anchor=south west,outer sep=0pt] 
    {\pgfimage[width=0.161\textwidth]{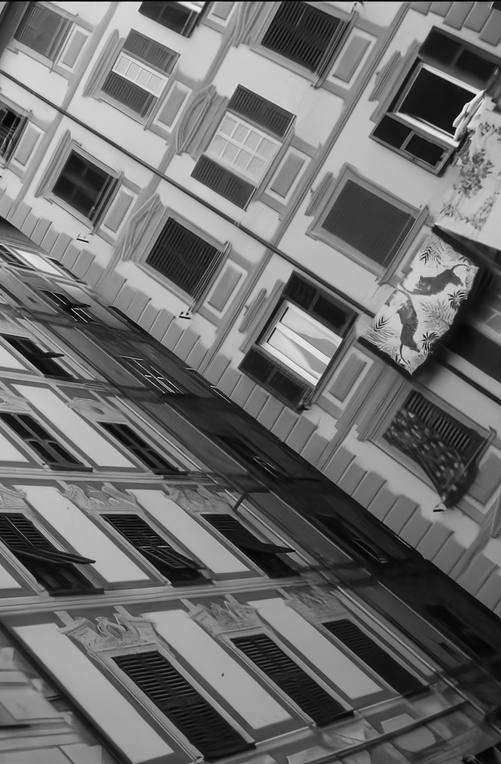}};
    {\spy[red!70!black] on (0.7,0.7) in node at (1.00,3.51);}
    \end{tikzpicture}
}
\caption{Extract from Urban100 scene 13, $\sigma=25$. Left to right: ground truth, noisy ($20.16$ dB), BM3D ($30.40$ dB), DnCNN ($30.71$ dB), NLRN ($31.41$ dB), GCDN (\textbf{31.53 dB}).}
\label{fig:natural_visual}
\end{figure*}

\begin{figure*}
\vspace*{3.2cm}
\centering
\parbox{0.195\textwidth}{%
\begin{tikzpicture}[overlay,remember picture,spy using outlines={rectangle,magnification=3,size=1cm}]
    \node[inner sep=0pt, anchor=south west,outer sep=0pt] 
    {\pgfimage[width=0.195\textwidth]{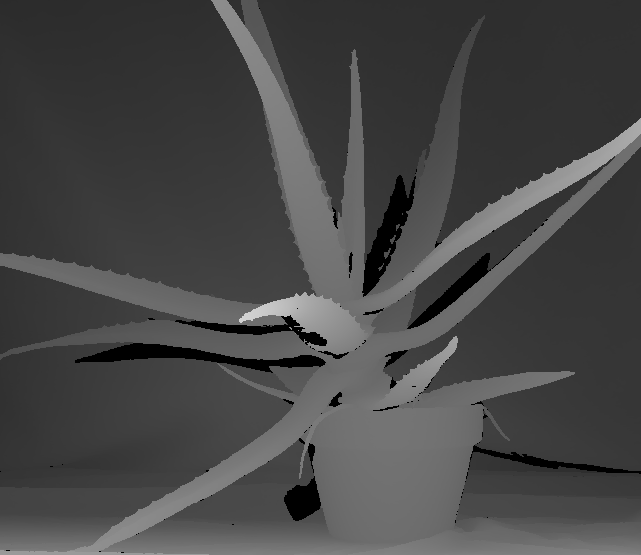}};
    {\spy[red!70!black] on (1.70,1.47) in node at (3.08,0.51);}
\end{tikzpicture}
}
\parbox{0.195\textwidth}{%
\begin{tikzpicture}[overlay,remember picture,spy using outlines={rectangle,magnification=3,size=1cm}]
    \node[inner sep=0pt, anchor=south west,outer sep=0pt] 
    {\pgfimage[width=0.195\textwidth]{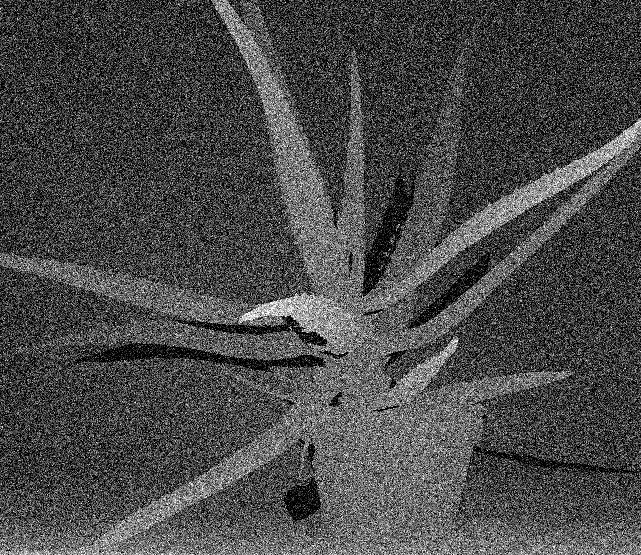}};
    {\spy[red!70!black] on (1.70,1.47) in node at (3.08,0.51);}
\end{tikzpicture}
}
\parbox{0.195\textwidth}{%
\begin{tikzpicture}[overlay,remember picture,spy using outlines={rectangle,magnification=3,size=1cm}]
    \node[inner sep=0pt, anchor=south west,outer sep=0pt] 
    {\pgfimage[width=0.195\textwidth]{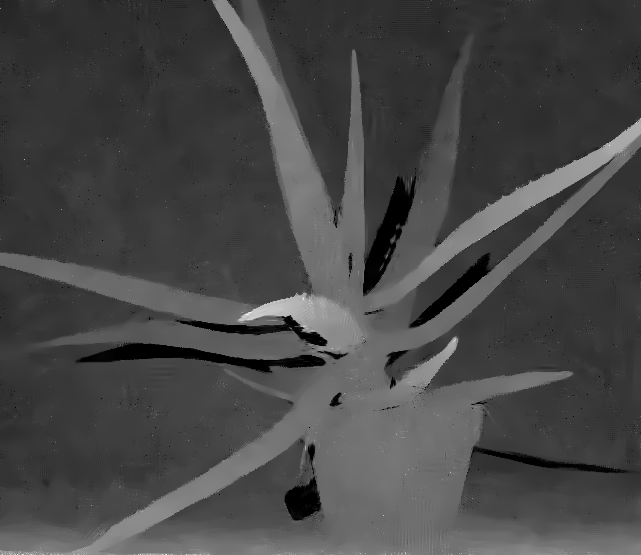}};
    {\spy[red!70!black] on (1.70,1.47) in node at (3.08,0.51);}
\end{tikzpicture}
}
\parbox{0.195\textwidth}{%
\begin{tikzpicture}[overlay,remember picture,spy using outlines={rectangle,magnification=3,size=1cm}]
    \node[inner sep=0pt, anchor=south west,outer sep=0pt] 
    {\pgfimage[width=0.195\textwidth]{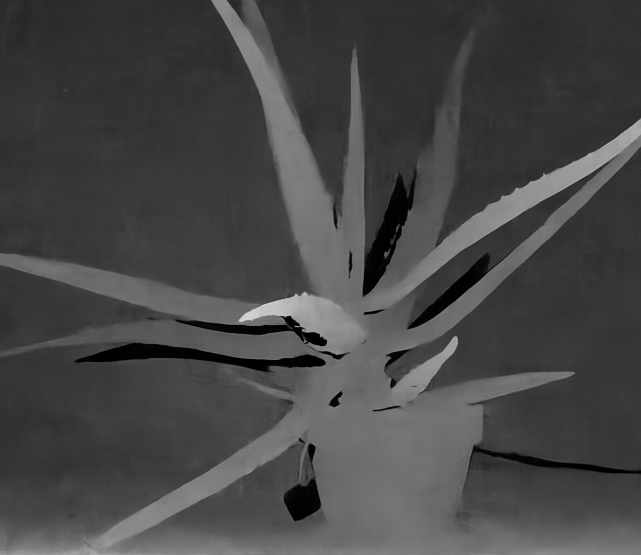}};
    {\spy[red!70!black] on (1.70,1.47) in node at (3.08,0.51);}
\end{tikzpicture}
}
\parbox{0.195\textwidth}{%
\begin{tikzpicture}[overlay,remember picture,spy using outlines={rectangle,magnification=3,size=1cm}]
    \node[inner sep=0pt, anchor=south west,outer sep=0pt] 
    {\pgfimage[width=0.195\textwidth]{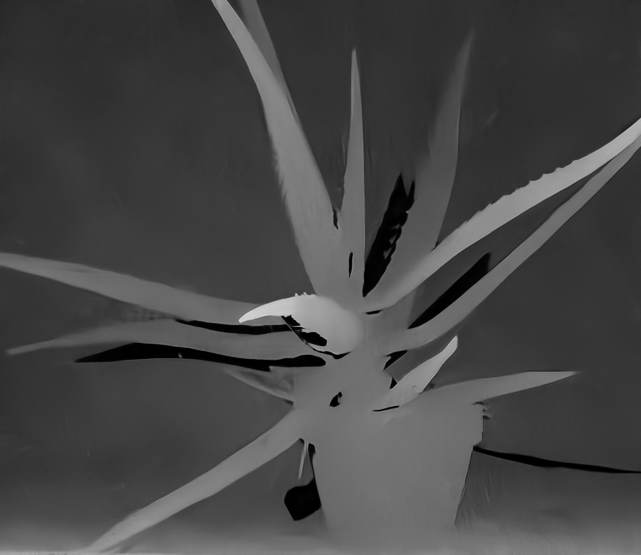}};
    {\spy[red!70!black] on (1.70,1.47) in node at (3.08,0.51);}
\end{tikzpicture}
}
\caption{\textit{aloe} depthmap denoising, $\sigma=50$. Left to right: ground truth, noisy ($14.16$ dB), OGLR ($32.24$ dB), NLRN ($33.23$ dB), GCDN (\textbf{33.37 dB}).}
\label{fig:depth_visual}
\end{figure*}

In this section we compare the proposed network with state-of-the-art models for the Gaussian denoising task of grayscale images. We train an independent model for each noise standard deviation, which is assumed to be known a priori for all methods. We fix the number of neighbors for the proposed method to 16. The reference methods can be classified into model-based algorithms such as BM3D \cite{dabov2007image}, WNNM \cite{gu2014weighted}, TNRD \cite{chen2016trainable} and recent deep-learning methods such as DnCNN \cite{zhang2017beyond}, N$^3$Net \cite{plotz2018neural} and NLRN \cite{liu2018non}. In particular, among the deep-learning methods, N$^3$Net and NLRN propose non-local approaches. All results have been obtained running the pretrained models provided by the authors, except for N$^3$Net at $\sigma=15$ which is unavailable. Table \ref{table:results} reports the PSNR and SSIM values obtained for the Set12, BSD68 and Urban100 standard test sets. It can be seen that the proposed method achieves state-of-the art performance and works especially well at low to medium levels of noise. This can be explained by a higher difficulty in constructing a meaningful graph from the noisy image at higher noise levels. We also notice that the proposed method achieves strong results on the Urban dataset. This dataset contains higher resolution images with respect to the other two and is mainly composed of photos of buildings and other regular structures where exploiting self-similarity is very important. In addition, it is also worth mentioning that the proposed method provides a better visual quality. In many cases, the proposed method has a higher SSIM score, even if NRLN has better performance in terms of PSNR. This can also be noticed in Fig. \ref{fig:natural_visual}, which shows a visual comparison on an image from the Urban100 dataset. In general, the images produced by the proposed algorithm present sharper edges and smoother content in uniform areas.
We can notice that many areas in the photos from Urban100 have approximately piecewise smooth characteristics. It is well known that image processing algorithms based on graphs are well suited for piecewise smooth content (see, e.g., \cite{Hu2015multiresolution,Fracastoro2017graph} in the context of compression and \cite{pang2017graph} for denoising). To further show this point, we study the performance of the proposed method for denoising of depth maps, e.g., generated by time-of-flight cameras. The OGLR algorithm \cite{pang2017graph} based on a graph smoothness regularizer achieved state-of-the-art results among model-based algorithms for this specific task where it is essential to preserve edge sharpness while simultaneously smoothing the flat areas. Table \ref{table:depthmap} reports the PSNR and SSIM results achieved on a standard set of depth maps\footnote{http://vision.middlebury.edu/stereo/data/.}. It can be seen that the proposed method outperforms both NLRN and OGLR, even at high levels of noise. Also, we can notice that OGLR displays competitive performance at low noise levels, but its visual quality significantly degrades when in presence of stronger noise. Fig. \ref{fig:depth_visual} shows a visual comparison where it can be seen that GCDN produces sharper edges while also providing a very smooth background.

\subsection{Real image denoising}
\begin{figure*}
\vspace*{3cm}
\centering
\parbox{0.161\textwidth}{%
\begin{tikzpicture}[overlay,remember picture,spy using outlines={rectangle,magnification=2,width=1.9cm, height=2.3cm}]
    \node[inner sep=0pt, anchor=south west,outer sep=0pt] 
    {\pgfimage[width=0.161\textwidth]{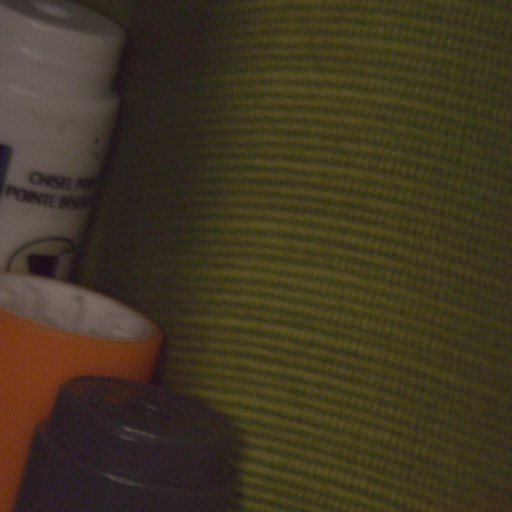}};
    {\spy[red!70!black] on (0.5,1.5) in node at (2,1.8);}
\end{tikzpicture}
}
\parbox{0.161\textwidth}{%
\begin{tikzpicture}[overlay,remember picture,spy using outlines={rectangle,magnification=2,width=1.9cm, height=2.3cm}]
    \node[inner sep=0pt, anchor=south west,outer sep=0pt] 
    {\pgfimage[width=0.161\textwidth]{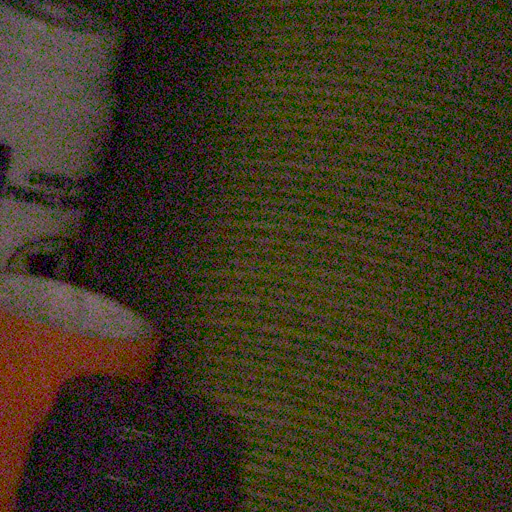}};
    {\spy[red!70!black] on (0.5,1.5) in node at (2,1.8);}
\end{tikzpicture}
}
\parbox{0.161\textwidth}{%
\begin{tikzpicture}[overlay,remember picture,spy using outlines={rectangle,magnification=2,width=1.9cm, height=2.3cm}]
    \node[inner sep=0pt, anchor=south west,outer sep=0pt] 
    {\pgfimage[width=0.161\textwidth]{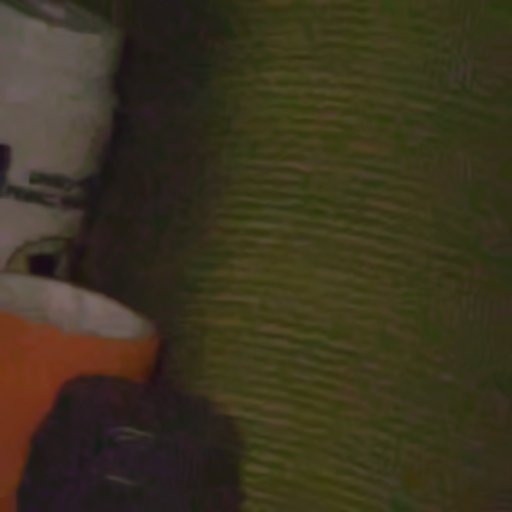}};
    {\spy[red!70!black] on (0.5,1.5) in node at (2,1.8);}
\end{tikzpicture}
}
\parbox{0.161\textwidth}{%
\begin{tikzpicture}[overlay,remember picture,spy using outlines={rectangle,magnification=2,width=1.9cm, height=2.3cm}]
    \node[inner sep=0pt, anchor=south west,outer sep=0pt] 
    {\pgfimage[width=0.161\textwidth]{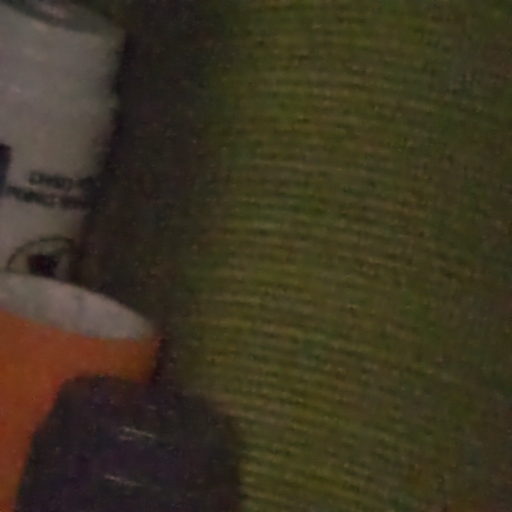}};
    {\spy[red!70!black] on (0.5,1.5) in node at (2,1.8);}
\end{tikzpicture}
}
\parbox{0.161\textwidth}{%
\begin{tikzpicture}[overlay,remember picture,spy using outlines={rectangle,magnification=2,width=1.9cm, height=2.3cm}]
    \node[inner sep=0pt, anchor=south west,outer sep=0pt] 
    {\pgfimage[width=0.161\textwidth]{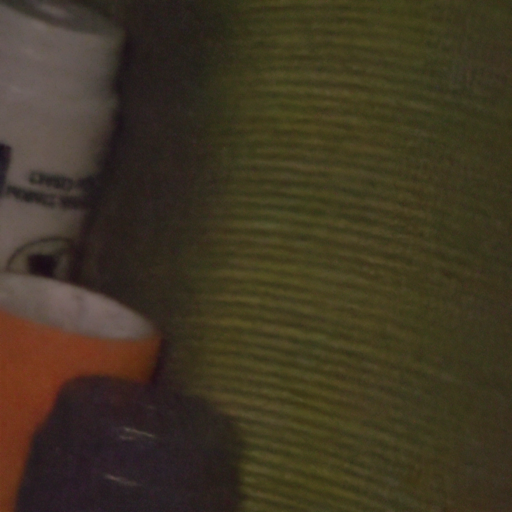}};
    {\spy[red!70!black] on (0.5,1.5) in node at (2,1.8);}
\end{tikzpicture}
}
\parbox{0.161\textwidth}{%
\begin{tikzpicture}[overlay,remember picture,spy using outlines={rectangle,magnification=2,width=1.9cm, height=2.3cm}]
    \node[inner sep=0pt, anchor=south west,outer sep=0pt] 
    {\pgfimage[width=0.161\textwidth]{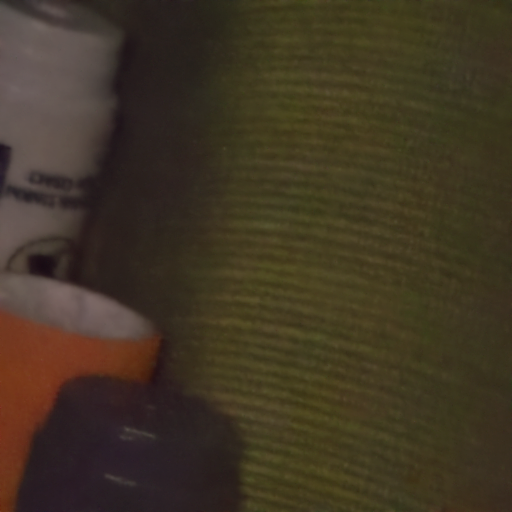}};
    {\spy[red!70!black] on (0.5,1.5) in node at (2,1.8);}
\end{tikzpicture}
}
\caption{Real image denoising. Left to right: ground truth, noisy (23.80 dB), CBM3D (34.84 dB), DnCNN (36.05 dB), NLRN (37.15 dB), GCDN (\textbf{37.33 dB}).}
\label{fig:real_visual}
\end{figure*}

Real image noise is generally more challenging than synthetic Gaussian noise. There are multiple contributions such as quantization noise, shot noise, fixed-pattern noise \cite{holst1998ccd,lukavs2006digital}, dark current, etc. that make it overall signal-dependent. It has been observed \cite{plotz2017benchmarking,SIDD_2018_CVPR} that deep learning methods trained on synthetic Gaussian noise perform poorly in presence of real noise. However, suitable retraining with real data generally improves their performance. In this section, we study the behavior of the proposed network in a blind denoising setting with real noisy images acquired by smartphones. We retrain the proposed method, NLRN and DnCNN on the SIDD dataset \cite{SIDD_2018_CVPR} composed of 30000 high-resolution images acquired by smartphone cameras at varying illumination and ISO levels. The authors provide clean and carefully registered ground truths for all the available scenes, so that it is possible to perform a supervised training. We create training and testing subsets from the sRGB images in the SIDD dataset by selecting a range of noise levels. Our training set is composed of 3500 crops of size $512 \times 512$ whose RMSE with respect to the ground truth is below 15. The testing set is composed of 25 random crops of size $512 \times 512$ with noise in the same range as the training set. Table \ref{table:real_noise} reports the results for CBM3D \cite{dabov2007color}, DnCNN, NLRN and the proposed GCDN. Notice that CBM3D is not a blind method, so we provide an estimate of the noise standard deviation, as computed by a noise estimation algorithm \cite{Chen2015Efficient}.
We can notice that the proposed method achieves better results and this is confirmed by the visual comparison in Fig. \ref{fig:real_visual}.

\begin{table}[]
\centering
\caption{Real image denoising (SIDD dataset)}
\begin{tabular}{ccccc}
\hline
     & CBM3D    & DnCNN    & NLRN     & GCDN          \\ \hline
PSNR & 38.73 dB & 39.98 dB & 41.24 dB & \textbf{41.48 dB} \\ \hline
SSIM & 0.9587 & 0.9605   & 0.9652   & \textbf{0.9697}   \\ \hline
\end{tabular}
\label{table:real_noise}
\end{table}

\vspace{-0.2cm}
\section{Conclusions} \label{sec:conclusions}
\vspace{-0.2cm}
In this paper, we presented a graph-convolutional neural network targeted for image denoising. The proposed  graph-convolutional layer allows to exploit both local and non-local similarities, resulting in an adaptive receptive field. 
We showed that the proposed architecture can outperform state-of-the-art denoising methods, achieving very strong results on piecewise smooth images. Finally, we have also considered a real image denoising setting, showing that the proposed method can provide a significant performance gain.
Future work will focus on extending the proposed architecture to other inverse problems, such as super-resolution \cite{dong2014learning,bordone2019deepsum}.

\bibliographystyle{IEEEtran}

\end{document}

%% file: commands.tex
\usepackage{algorithm, algorithmic}
\usepackage{graphicx} 
\usepackage{amsmath} 
\usepackage{amssymb}  
\usepackage{amsthm}
\usepackage{amsfonts}
\usepackage{dsfont}
\usepackage{color}
\usepackage{bm}
\usepackage{hyperref}
\usepackage{times,amsmath,epsfig}
\usepackage{cite}
\usepackage{tikz}
\usepackage{xcolor}
\usepackage[english]{babel}
\usepackage[latin1]{inputenc}
\usepackage{subfigure}
\usepackage{hhline}
\usepackage{mathtools}
\usepackage{multirow}

\newcommand{\nub}{\bm{\nu}}
\newcommand{\mub}{\bm{\mu}}
\newcommand{\Lambdab}{\bm{\Lambda}}

\newcommand{\Ab}{\mathbf{A}}
\newcommand{\Ib}{\mathbf{I}}
\newcommand{\Lb}{\mathbf{L}}
\newcommand{\Ub}{\mathbf{U}}
\newcommand{\Hb}{\mathbf{H}}
\newcommand{\Wb}{\mathbf{W}}

\newcommand{\nb}{\mathbf{n}}

\newcommand{\xb}{\mathbf{x}}
\newcommand{\yb}{\mathbf{y}}
\newcommand{\zb}{\mathbf{z}}

\newcommand{\wb}{\mathbf{w}}

\newcommand{\fun}{\mathcal{F}}
\newcommand{\V}{\mathcal{V}}

\newcommand{\E}{\mathcal{E}}
\newcommand{\G}{\mathcal{G}}

\newcommand{\Nc}{\mathcal{N}}

\newcommand{\Sx}{\mathcal{S}}

\newcommand{\RR}{\mathbb{R}}

\newcommand{\argmin}[1]{\underset{#1}{\mathrm{argmin\,}}}